

%
%


\def\famname{
 \textfont0=\textrm \scriptfont0=\scriptrm
 \scriptscriptfont0=\sscriptrm
 \textfont1=\textmi \scriptfont1=\scriptmi
 \scriptscriptfont1=\sscriptmi
 \textfont2=\textsy \scriptfont2=\scriptsy \scriptscriptfont2=\sscriptsy
 \textfont3=\textex \scriptfont3=\textex \scriptscriptfont3=\textex
 \textfont4=\textbf \scriptfont4=\scriptbf \scriptscriptfont4=\sscriptbf
 \skewchar\textmi='177 \skewchar\scriptmi='177
 \skewchar\sscriptmi='177
 \skewchar\textsy='60 \skewchar\scriptsy='60
 \skewchar\sscriptsy='60
 \def\rm{\fam0 \textrm} \def\bf{\fam4 \textbf}}
\def\sca#1{scaled\magstep#1} \def\scah{scaled\magstephalf} 
\def\twelvepoint{
 \font\textrm=cmr12 \font\scriptrm=cmr8 \font\sscriptrm=cmr6
 \font\textmi=cmmi12 \font\scriptmi=cmmi8 \font\sscriptmi=cmmi6 
 \font\textsy=cmsy10 \sca1 \font\scriptsy=cmsy8
 \font\sscriptsy=cmsy6
 \font\textex=cmex10 \sca1
 \font\textbf=cmbx12 \font\scriptbf=cmbx8 \font\sscriptbf=cmbx6
 \font\it=cmti12
 \font\sectfont=cmbx12 \sca1
 \font\sectmath=cmmib10 \sca2
 \font\sectsymb=cmbsy10 \sca2
 \font\refrm=cmr10 \scah \font\refit=cmti10 \scah
 \font\refbf=cmbx10 \scah
 \def\twelverm{\textrm} \def\twelveit{\it} \def\twelvebf{\textbf}
 \famname \textrm 
 \advance\voffset by .06in \advance\hoffset by .28in
 \normalbaselineskip=17.5pt plus 1pt \baselineskip=\normalbaselineskip
 \parindent=21pt
 \setbox\strutbox=\hbox{\vrule height10.5pt depth4pt width0pt}}


\catcode`@=11

{\catcode`\'=\active \def'{{}^\bgroup\prim@s}}

\def\screwcount{\alloc@0\count\countdef\insc@unt}   
\def\screwdimen{\alloc@1\dimen\dimendef\insc@unt} 
\def\screwbox{\alloc@4\box\chardef\insc@unt}

\catcode`@=12


\overfullrule=0pt			
\vsize=9in \hsize=6in
\lineskip=0pt				
\abovedisplayskip=1.2em plus.3em minus.9em 
\belowdisplayskip=1.2em plus.3em minus.9em	
\abovedisplayshortskip=0em plus.3em	
\belowdisplayshortskip=.7em plus.3em minus.4em	
\parindent=21pt
\setbox\strutbox=\hbox{\vrule height10.5pt depth4pt width0pt}
\def\makefootline{\baselineskip=30pt \line{\the\footline}}
\footline={\ifnum\count0=1 \hfil \else\hss\twelverm\folio\hss \fi}
\pageno=1


\def\put(#1,#2)#3{\screwdimen\unit  \unit=1in
	\vbox to0pt{\kern-#2\unit\hbox{\kern#1\unit
	\vbox{#3}}\vss}\nointerlineskip}


\def\\{\hfil\break}
\def\newpage{\vfill\eject}
\def\center{\leftskip=0pt plus 1fill \rightskip=\leftskip \parindent=0pt
 \def\textindent##1{\par\hangindent21pt\footrm\noindent\hskip21pt
 \llap{##1\enspace}\ignorespaces}\par}
\def\unnarrower{\leftskip=0pt \rightskip=\leftskip}


\def\vol#1 {{\refbf#1} }		 


\def\NP #1 {{\refit Nucl. Phys.} {\refbf B{#1}} }
\def\PL #1 {{\refit Phys. Lett.} {\refbf{#1}} }
\def\PR #1 {{\refit Phys. Rev. Lett.} {\refbf{#1}} }
\def\PRD #1 {{\refit Phys. Rev.} {\refbf D{#1}} }


\hyphenation{pre-print}
\hyphenation{quan-ti-za-tion}

%
%


\def\oonoo#1#2#3{\vbox{\ialign{##\crcr
	\hfil\hfil\hfil{$#3{#1}$}\hfil\crcr\noalign{\kern1pt\nointerlineskip}
	$#3{#2}$\crcr}}}
\def\oon#1#2{\mathchoice{\oonoo{#1}{#2}{\displaystyle}}
	{\oonoo{#1}{#2}{\textstyle}}{\oonoo{#1}{#2}{\scriptstyle}}
	{\oonoo{#1}{#2}{\scriptscriptstyle}}}
\def\dt#1{\oon{\hbox{\bf .}}{#1}}  
\def\ddt#1{\oon{\hbox{\bf .\kern-1pt.}}#1}    
\def\slap#1#2{\setbox0=\hbox{$#1{#2}$}
	#2\kern-\wd0{\hfuzz=1pt\hbox to\wd0{\hfil$#1{/}$\hfil}}}
\def\sla#1{\mathpalette\slap{#1}}                
\def\bop#1{\setbox0=\hbox{$#1M$}\mkern1.5mu
	\lower.02\ht0\vbox{\hrule height0pt depth.06\ht0
	\hbox{\vrule width.06\ht0 height.9\ht0 \kern.9\ht0
	\vrule width.06\ht0}\hrule height.06\ht0}\mkern1.5mu}
\def\bo{{\mathpalette\bop{}}}                        
\def~{\widetilde} 
\mathcode`\*="702A                  
\def\in{\relax\ifmmode\mathchar"3232\else{\refit in\/}\fi} 
\def\f#1#2{{\textstyle{#1\over#2}}}	   
\def\half{{\textstyle{1\over{\raise.1ex\hbox{$\scriptstyle{2}$}}}}}

\def\Gamma{\mathchar"0100}
\def\Delta{\mathchar"0101}
\def\Theta{\mathchar"0102}
\def\Lambda{\mathchar"0103}
\def\Xi{\mathchar"0104}
\def\Pi{\mathchar"0105}
\def\Sigma{\mathchar"0106}
\def\Upsilon{\mathchar"0107}
\def\Phi{\mathchar"0108}
\def\Psi{\mathchar"0109}
\def\Omega{\mathchar"010A}

\catcode128=13 \def €{\"A}                 
\catcode129=13 \def {\AA}                 
\catcode130=13 \def '{\c}           	   
\catcode131=13 \def ƒ{\'E}                   
\catcode132=13 \def "{\~N}                   
\catcode133=13 \def …{\"O}                 
\catcode134=13 \def †{\"U}                  
\catcode135=13 \def ‡{\'a}                  
\catcode136=13 \def ˆ{\`a}                   
\catcode137=13 \def ‰{\^a}                 
\catcode138=13 \def Š{\"a}                 
\catcode139=13 \def ‹{\~a}                   
\catcode140=13 \def Œ{\alpha}            
\catcode141=13 \def {\chi}                
\catcode142=13 \def Ž{\'e}                   
\catcode143=13 \def {\`e}                    
\catcode144=13 \def {\^e}                  
\catcode145=13 \def '{\"e}                
\catcode146=13 \def '{\'\i}                 
\catcode147=13 \def "{\`\i}                  
\catcode148=13 \def "{\^\i}                
\catcode149=13 \def •{\"\i}                
\catcode150=13 \def –{\~n}                  
\catcode151=13 \def —{\'o}                 
\catcode152=13 \def ˜{\`o}                  
\catcode153=13 \def ™{\^o}                
\catcode154=13 \def š{\"o}                 
\catcode155=13 \def ›{\~o}                  
\catcode156=13 \def œ{\'u}                  
\catcode157=13 \def {\`u}                  
\catcode158=13 \def ž{\^u}                
\catcode159=13 \def Ÿ{\"u}                
\catcode160=13 \def  {\tau}               
\catcode161=13 \mathchardef ¡="2203     
\catcode162=13 \def ¢{\oplus}           
\catcode163=13 \def £{\relax\ifmmode\to\else\itemize\fi} 
\catcode164=13 \def ¤{\subset}	  
\catcode165=13 \def ¥{\infty}           
\catcode166=13 \def ¦{\mp}                
\catcode167=13 \def §{\sigma}           
\catcode168=13 \def ¨{\rho}               
\catcode169=13 \def ©{\gamma}         
\catcode170=13 \def ª{\leftrightarrow} 
\catcode171=13 \def «{\relax\ifmmode\acute\else\expandafter\'\fi}
\catcode172=13 \def ¬{\relax\ifmmode\expandafter\ddt\else\expandafter\"\fi}
\catcode173=13 \def ­{\equiv}            
\catcode174=13 \def ®{\approx}          
\catcode175=13 \def ¯{\Omega}          
\catcode176=13 \def °{\otimes}          
\catcode177=13 \def ±{\ne}                 
\catcode178=13 \def ²{\le}                   
\catcode179=13 \def ³{\ge}                  
\catcode180=13 \def ´{\upsilon}          
\catcode181=13 \def µ{\mu}                
\catcode182=13 \def ¶{\delta}             
\catcode183=13 \def ·{\epsilon}          
\catcode184=13 \def ¸{\Pi}                  
\catcode185=13 \def ¹{\pi}                  
\catcode186=13 \def º{\beta}               
\catcode187=13 \def »{\partial}           
\catcode188=13 \def ¼{\nobreak\ }       
\catcode189=13 \def ½{\zeta}               
\catcode190=13 \def ¾{\sim}                 
\catcode191=13 \def ¿{\omega}           
\catcode192=13 \def À{\dt}                     
\catcode193=13 \def Á{\gets}                
\catcode194=13 \def Â{\lambda}           
\catcode195=13 \def Ã{\nu}                   
\catcode196=13 \def Ä{\phi}                  
\catcode197=13 \def Å{\xi}                     
\catcode198=13 \def Æ{\psi}                  
\catcode199=13 \def Ç{\int}                    
\catcode200=13 \def È{\oint}                 
\catcode201=13 \def É{\relax\ifmmode\cdot\else\vol\fi}    
\catcode202=13 \def Ê{\relax\ifmmode\,\else\thinspace\fi}
\catcode203=13 \def Ë{\`A}                      
\catcode204=13 \def Ì{\~A}                      
\catcode205=13 \def Í{\~O}                      
\catcode206=13 \def Î{\Theta}              
\catcode207=13 \def Ï{\theta}               
\catcode208=13 \def Ð{\relax\ifmmode\bar\else\expandafter\=\fi}
\catcode209=13 \def Ñ{\overline}             
\catcode210=13 \def Ò{\langle}               
\catcode211=13 \def Ó{\relax\ifmmode\{\else\ital\fi}      
\catcode212=13 \def Ô{\rangle}               
\catcode213=13 \def Õ{\}}                        
\catcode214=13 \def Ö{\sla}                      
\catcode215=13 \def ×{\relax\ifmmode\check\else\expandafter\v\fi}
\catcode216=13 \def Ø{\"y}                     
\catcode217=13 \def Ù{\"Y}  		    
\catcode218=13 \def Ú{\Leftarrow}       
\catcode219=13 \def Û{\Leftrightarrow}       
\catcode220=13 \def Ü{\relax\ifmmode\Rightarrow\else\sect\fi}
\catcode221=13 \def Ý{\sum}                  
\catcode222=13 \def Þ{\prod}                 
\catcode223=13 \def ß{\widehat}              
\catcode224=13 \def à{\pm}                     
\catcode225=13 \def á{\nabla}                
\catcode226=13 \def â{\quad}                 
\catcode227=13 \def ã{\in}               	
\catcode228=13 \def ä{\star}      	      
\catcode229=13 \def å{\sqrt}                   
\catcode230=13 \def æ{\^E}			
\catcode231=13 \def ç{\Upsilon}              
\catcode232=13 \def è{\"E}    	   	 
\catcode233=13 \def é{\`E}               	  
\catcode234=13 \def ê{\Sigma}                
\catcode235=13 \def ë{\Delta}                 
\catcode236=13 \def ì{\Phi}                     
\catcode237=13 \def í{\`I}        		   
\catcode238=13 \def î{\iota}        	     
\catcode239=13 \def ï{\Psi}                     
\catcode240=13 \def ð{\times}                  
\catcode241=13 \def ñ{\Lambda}             
\catcode242=13 \def ò{\cdots}                
\catcode243=13 \def ó{\^U}			
\catcode244=13 \def ô{\`U}    	              
\catcode245=13 \def õ{\bo}                       
\catcode246=13 \def ö{\relax\ifmmode\hat\else\expandafter\^\fi}
\catcode247=13 \def÷{\relax\ifmmode\tilde\else\expandafter\~\fi}
\catcode248=13 \def ø{\ll}                         
\catcode249=13 \def ù{\gg}                       
\catcode250=13 \def ú{\eta}                      
\catcode251=13 \def û{\kappa}                  
\catcode252=13 \def ü{\half}     		 
\catcode253=13 \def ý{\Gamma} 		
\catcode254=13 \def þ{\Xi}   			
\catcode255=13 \def ÿ{\relax\ifmmode{}^{\dagger}{}\else\dag\fi}


\def\ital#1Õ{{\it#1\/}}	     
\def\un#1{\relax\ifmmode\underline#1\else $\underline{\hbox{#1}}$
	\relax\fi}

\def\roonoo#1#2#3{\vbox{\ialign{##\crcr
	\hfil{$#3{#1}$}\hfil\crcr\noalign{\kern1pt\nointerlineskip}
	$#3{#2}$\crcr}}}
\def\roon#1#2{\mathchoice{\roonoo{#1}{#2}{\displaystyle}}
	{\roonoo{#1}{#2}{\textstyle}}{\roonoo{#1}{#2}{\scriptstyle}}
	{\roonoo{#1}{#2}{\scriptscriptstyle}}}
\def\rdt#1{\roon{\hbox{\bf .}}{#1}}  
\def\tdt#1{\oon{\hbox{\bf .\kern-1pt.\kern-1pt.}}#1}   
\def\({\eqno(}
\def\li{\openup1\jot \eqalignno}


\def\õ#1{
	\screwcount\num
	\num=1
	\screwdimen\downsy
	\downsy=-1.5ex
	\mkern-3.5mu
	õ
	\loop
	\ifnum\num<#1
	\llap{\raise\num\downsy\hbox{$õ$}}
	\advance\num by1
	\repeat}
\def\upõ#1#2{\screwcount\numup
	\numup=#1
	\advance\numup by-1
	\screwdimen\upsy
	\upsy=.75ex
	\mkern3.5mu
	\raise\numup\upsy\hbox{$#2$}}



\newcount\marknumber	\marknumber=1
\newcount\countdp \newcount\countwd \newcount\countht 

%
%
\ifx\pdfoutput\undefined
\def\rgboo#1{}
\input epsf

\def\postscript#1{\special{" #1}}		
\postscript{
	/bd {bind def} bind def
	/fsd {findfont exch scalefont def} bd
	/sms {setfont moveto show} bd
	/ms {moveto show} bd
	/pdfmark where		
	{pop} {userdict /pdfmark /cleartomark load put} ifelse
	[ /PageMode /UseOutlines		
	/DOCVIEW pdfmark}
\def\bookmark#1#2{\postscript{		
	[ /Dest /MyDest\the\marknumber /View [ /XYZ null null null ] /DEST pdfmark
	[ /Title (#2) /Count #1 /Dest /MyDest\the\marknumber /OUT pdfmark}%
	\advance\marknumber by1}
\def\pdfklink#1#2{%
	\hskip-.25em\setbox0=\hbox{#1}%
		\countdp=\dp0 \countwd=\wd0 \countht=\ht0%
		\divide\countdp by65536 \divide\countwd by65536%
			\divide\countht by65536%
		\advance\countdp by1 \advance\countwd by1%
			\advance\countht by1%
		\def\linkdp{\the\countdp} \def\linkwd{\the\countwd}%
			\def\linkht{\the\countht}%
	\postscript{
		[ /Rect [ -1.5 -\linkdp.0 0\linkwd.0 0\linkht.5 ] 
		/Border [ 0 0 0 ]
		/Action << /Subtype /URI /URI (#2) >>
		/Subtype /Link
		/ANN pdfmark}{\rgb{1 0 0}{#1}}}
%
%
\else
\def\rgboo#1{\pdfliteral{#1 rg #1 RG}}

\pdfcatalog{/PageMode /UseOutlines}		
\def\bookmark#1#2{
	\pdfdest num \marknumber xyz
	\pdfoutline goto num \marknumber count #1 {#2}
	\advance\marknumber by1}
\def\pdfklink#1#2{%
	\noindent\pdfstartlink user
		{/Subtype /Link
		/Border [ 0 0 0 ]
		/A << /S /URI /URI (#2) >>}{\rgb{1 0 0}{#1}}%
	\pdfendlink}
\fi

\def\rgbo#1#2{\rgboo{#1}#2\rgboo{0 0 0}}
\def\rgb#1#2{\mark{#1}\rgbo{#1}{#2}\mark{0 0 0}}
\def\pdflink#1{\pdfklink{#1}{#1}}
\def\xxxlink#1{\pdfklink{[arXiv:#1]}{http://arXiv.org/abs/#1}}

\catcode`@=11

\def\wlog#1{}	


\def\makeheadline{\vbox to\z@{\vskip-36.5\p@
	\line{\vbox to8.5\p@{}\the\headline%
	\ifnum\pageno=\z@\rgboo{0 0 0}\else\rgboo{\topmark}\fi%
	}\vss}\nointerlineskip}
\headline={
	\ifnum\pageno=\z@
		\hfil
	\else
		\ifnum\pageno<\z@
			\ifodd\pageno
				\tenrm\romannumeral-\pageno\hfil\lefthead\hfil
			\else
				\tenrm\hfil\righthead\hfil\romannumeral-\pageno
			\fi
		\else
			\ifodd\pageno
				\tenrm\hfil\righthead\hfil\number\pageno
			\else
				\tenrm\number\pageno\hfil\lefthead\hfil
			\fi
		\fi
	\fi}

\catcode`@=12

\def\righthead{\hfil} \def\lefthead{\hfil}
\nopagenumbers


\def\chrulefill{\rgb{1 0 0}{\hrulefill}}
\def\cdotfill{\rgb{1 0 0}{\dotfill}}
\newcount\area	\area=1
\newcount\cross	\cross=1
\def\volume#1\par{\newpage\noindent{\biggest{\rgb{1 .5 0}{#1}}}
	\par\nobreak\bigskip\medskip\area=0}
\def\chapskip{\par\ifnum\area=0\bigskip\medskip\goodbreak
	\else\newpage\fi}
\def\chapy#1{\area=1\cross=0
	\xdef\lefthead{\rgbo{1 0 .5}{#1}}\vbox{\biggerer\offinterlineskip
	\line{\chrulefill¼\hphantom{\lefthead}\chrulefill}
	\line{\chrulefill¼\lefthead\chrulefill}}\par\nobreak\medskip}
\def\chap#1\par{\chapskip\bookmark3{#1}\chapy{#1}}
\def\sectskip{\par\ifnum\cross=0\bigskip\medskip\goodbreak
	\else\newpage\fi}
\def\secty#1{\cross=1
	\xdef\righthead{\rgbo{1 0 1}{#1}}\vbox{\bigger\offinterlineskip
	\line{\cdotfill¼\hphantom{\righthead}\cdotfill}
	\line{\cdotfill¼\righthead\cdotfill}}\par\nobreak\medskip}
\def\sect#1 #2\par{\sectskip\bookmark{#1}{#2}\secty{#2}}
\def\subsectskip{\par\ifdim\lastskip<\medskipamount
	\bigskip\medskip\goodbreak\else\nobreak\fi}
\def\subsecty#1{\noindent{\sectfont{\rgbo{.5 0 1}{#1}}}\par\nobreak\medskip}
\def\subsect#1\par{\subsectskip\bookmark0{#1}\subsecty{#1}}
\long\def\x#1 #2\par{\hangindent2\parindent%
\mark{0 0 1}\rgboo{0 0 1}{\bf Exercise #1}\\#2%
\par\rgboo{0 0 0}\mark{0 0 0}}
\def\refs{\bigskip\noindent{\bf \rgbo{0 .5 1}{REFERENCES}}\par\nobreak\medskip
	\frenchspacing \parskip=0pt \refrm \baselineskip=1.23em plus 1pt
	\def\ital##1Õ{{\refit##1\/}}}
\long\def\twocolumn#1#2{\hbox to\hsize{\vtop{\hsize=2.9in#1}
	\hfil\vtop{\hsize=2.9in #2}}}


\twelvepoint
\font\bigger=cmbx12 \sca2
\font\biggerer=cmb10 \sca5
\font\biggest=cmssdc10 scaled 3700
 \sca5

 \sca3


\def Ü{\relax\ifmmode\Rightarrow\else\expandafter\subsect\fi}
\def Û{\relax\ifmmode\Leftrightarrow\else\expandafter\sect\fi}
\def Ú{\relax\ifmmode\Leftarrow\else\expandafter\chap\fi}

\def\itemize#1 {\item{\bf#1}}
\def\itemizze#1 {\itemitem{\bf#1}}
\def\itemutem{\par\indent\indent \hangindent3\parindent \textindent}
\def\itemizzze#1 {\itemutem{\bf#1}}
\def ª{\relax\ifmmode\leftrightarrow\else\itemizze\fi}
\def Á{\relax\ifmmode\gets\else\itemizzze\fi}

\def\¢{\ominus}
\def\A{{\cal A}}  \def\B{{\cal B}}  \def\C{{\cal C}}  
    \def\G{{\cal G}}  \def\H{{\cal H}}   
      \def\M{{\cal M}}     
\def\O{{\cal O}}

\def\Ä{\varphi}  \def\¿{\varpi}	\def\Ï{\vartheta}

\def ò{\relax\ifmmode\cdots\else\dotfill\fi}

\chardef\slo="1C


\def\cvrule{\rgbo{0 .5 1}{\vrule}}
\def\chrule{\rgbo{0 .5 1}{\hrule}}
\def\boxit#1{\leavevmode\thinspace\hbox{\cvrule\vtop{\vbox{\chrule%
	\vskip3pt\kern1pt\hbox{\vphantom{\bf/}\thinspace\thinspace%
	{\bf#1}\thinspace\thinspace}}\kern1pt\vskip3pt\chrule}\cvrule}%
	\thinspace}
\def\Boxit#1{\noindent\vbox{\chrule\hbox{\cvrule\kern3pt\vbox{
	\advance\hsize-7pt\vskip-\parskip\kern3pt\bf#1
	\hbox{\vrule height0pt depth\dp\strutbox width0pt}
	\kern3pt}\kern3pt\cvrule}\chrule}}




\def\today{\ifcase\month\or
 January\or February\or March\or April\or May\or June\or July\or
 August\or September\or October\or November\or December\fi
 \space\number\day, \number\year}

\parindent=20pt
\newskip\normalparskip	\normalparskip=.7\medskipamount
\parskip=\normalparskip	



\catcode`\|=\active \catcode`\<=\active \catcode`\>=\active 
\def|{\relax\ifmmode\delimiter"026A30C \else$\mathchar"026A$\fi}
\def<{\relax\ifmmode\mathchar"313C \else$\mathchar"313C$\fi}
\def>{\relax\ifmmode\mathchar"313E \else$\mathchar"313E$\fi}


%
%
%
%
%
%
%

\def\thetitle#1#2#3#4#5{
 \def\titlefont{\biggest} \font\footrm=cmr10 \font\footit=cmti10
  \twelverm
	{\hbox to\hsize{#4 \hfill YITP-SB-#3}}\par
	\vskip.8in minus.1in {\center\baselineskip=2.2\normalbaselineskip
 {\titlefont #1}\par}{\center\baselineskip=\normalbaselineskip
 \vskip.5in minus.2in #2
	\vskip1.4in minus1.2in {\twelvebf ABSTRACT}\par}
 \vskip.1in\par
 \narrower\par#5\par\unnarrower\vskip3.5in minus3.3in\eject}
\def\paper\par#1\par#2\par#3\par#4\par#5\par{
	\thetitle{#1}{#2}{#3}{#4}{#5}} 
\def\author#1#2{#1 \vskip.1in {\twelveit #2}\vskip.1in}
\def\YITP{C. N. Yang Institute for Theoretical Physics\\
	State University of New York, Stony Brook, NY 11794-3840}
\def\WS{W. Siegel\footnote{$*$}{
	\pdflink{mailto:siegel@insti.physics.sunysb.edu}\\
	\pdfklink{http://insti.physics.sunysb.edu/\~{}siegel/plan.html}
	{http://insti.physics.sunysb.edu/\noexpand~siegel/plan.html}}}


\def\back#1{\rgbo{#1}{\llap{\vrule height12pt depth5.5pt width.5em}\leaders\hrule height12pt depth5.5pt\hfill\rlap{\vrule height12pt depth5.5pt width.5em}}}

\def\backt#1{\rgbo{#1}{\llap{\vrule height12pt depth-3.25pt width.5em}\leaders\hrule height12pt depth-3.25pt\hfill\rlap{\vrule height12pt depth-3.25pt width.5em}}}

\def\backb#1{\rgbo{#1}{\llap{\vrule height3.25pt depth5.5pt width.5em}\leaders\hrule height3.25pt depth5.5pt\hfill\rlap{\vrule height3.25pt depth5.5pt width.5em}}}

\pageno=0

\paper

{\rgb{1 0 0.33}{New superspaces/algebras\\ for superparticles/strings
}}

\author\WS\YITP

11-18

June 8, 2011

We describe covariant derivatives with respect to the coordinates of the full superPoincar«e group and dual coordinates, for Yang-Mills and supergravity.  The derivatives have engineering dimension running from 0 to 2.  Prepotentials appear as potentials for Lorentz derivatives (spin).  Their role is clarified in a lightcone analysis, where they also act as compensating gauge parameters and Hertz potentials.  Field strengths appear as potentials for dual-coordinate derivatives, until dimensional reduction.  These generalizations extend the superstring's affine Lie algebra, and generalize gauge couplings for the superparticle.

\pageno=2

Û0 Introduction

Motivated by constructions of currents for first-quantization of superparticles and superstrings, we extend the covariant derivatives of (super) Yang-Mills and gravity to include not only those corresponding to translations, spin, and (for the supersymmetric cases) supersymmetry, but also those ``dual" to spin and supersymmetry.  There are extra coordinates associated with the extra derivatives, but they can be eliminated by both (1) isotropy (coset) constraints, for the gauge-covariant derivative for spin, and (2) dimensional reduction, by constraining symmetry generators dual to spin and supersymmetry.  The dimensional reduction allows the usual field strengths to be interpreted as gauge fields of the extra dimensions, and exhibits a duality between field strengths and (pre)potentials.  The coset constraints define spin as a differential operator, so in gravity the vierbein and Lorentz connection can be treated on equal footing.  The extra coordinates allow prepotentials to appear as (undifferentiated) potentials in the covariant derivatives that otherwise would be of too low (engineering) dimension to do so.  R-symmetry can be treated in the same way as spin, and arises from spin upon (the usual kind of) dimensional reduction; ``$à$" directions in lightcone treatments can also be interpreted as directions in R-space.

In the next section we outline the general construction, introducing extra gauge fields by gauge covariantizing before eliminating extra coordinates.  From a first-quantization point of view, the extra derivatives are natural for coupling external fields, and their currents automatically arise in the action.  

In the following section we give some examples:  In N=0 Yang-Mills, gauge covariantizing the spin introduces a nontrivial dimensionless potential in the lightcone gauge, where it acts as both Hertz potential and compensating gauge parameter that automatically makes all Lorentz transformations gauge preserving.  In selfdual (4D) Yang-Mills (supersymmetric or not), the usual prepotential appears as this potential, automatically in the covariant derivatives without solving differential constraints.  In selfdual (gauged) supergravity, the dual derivatives come into play, allowing the usual prepotential, which now has negative dimension, to again appear as a potential (without derivatives acting on it) in the covariant spin derivative, as a ``connection" for the derivative dual to spin.  Similar remarks apply to the non-selfdual case.

We devote the subsequent section to a lightcone analysis of 10D super Yang-Mills.  The usual lightcone treatment can be derived straightforwardly from the covariant one, and suggests a covariant analysis for 4D N=4 when lightlike directions are treated as null, complex, spacelike directions in R-space.  Directions for further research are sugested in the conclusions.

Û4 First-quantization

ÜCosets

Superspace is generally defined as the coset space ($G/H$) of the symmetry (isometry) group superPoincar«e ($G$) mod the gauge (isotropy) group Lorentz ($H$).  As in the usual coset construction, the coordinates of the space parametrize a group element, with the global symmetry derived from group multiplication from one side and covariant derivatives from group multiplication on the other, converted into differential operators on the group space:
$$ ¶g(z) = (i·^A T_A) g + g (i½^A T_A) = (·^A q_A + ½^A d_A) g,ââ
	[T_A,T_BÕ = -if_{AB}{}^C T_C,ââ $$
$$ g^{-1}dg = iÊdz^M R_M{}^A T_A,ââ(dg)g^{-1} = iÊdz^M L_M{}^A T_A $$
$$ d_A = R_A{}^M »_M,ââq_A = L_A{}^M »_M $$
$$ [d_A,d_BÕ = f_{AB}{}^C d_C,ââ[q_A,q_BÕ = -f_{AB}{}^C q_C $$
where $g$ is a group element in a matrix (or Hilbert space) representation where the generators are represented by $T_A$, parametrized by coordinates $z^M$, $d_A$ are the covariant derivatives, $q_A$ are the symmetry generators, and $f_{AB}{}^C$ are the structure constants, the free value of the torsion.  For (compact) internal symmetries, the distinction between left and right multiplication is arbitrary; but for spacetime symmetries (especially supersymmetry) there are physical differences, as we'll explain in this subsection and the following one.

The representation space $|ÆÔ$ of the coset is then expressed as wave functions $Æ_i(z)­Òz,{}_i|ÆÔ$ with basis element $|z,{}^iÔ$ defined by
$$ |z,{}^iÔ ­ g(z)|0,{}^iÔ,ââT_{(H),I} |0,{}^iÔ = |0,{}^jÔH_{Ij}{}^iâÛâ
	d_{(H),I} Æ_i(z) = H_{Ii}{}^j Æ_j(z) $$
Only the derivatives $d_{(G/H)}$ act nontrivially on the coset, the other (isotropy) derivatives $d_{(H)}$ having been used as constraints (set equal to a matrix representation $H_I$ on the fields) to define it.  

In (super) Yang-Mills these symmetry-covariant derivatives are then also covariantized with respect to the Yang-Mills gauge group:
$$ d_A £ á_A = d_A + iA_A $$
$$ [á_A,á_BÕ = f_{AB}{}^C á_C +iF_{AB} $$
(The Yang-Mills gauge group is unrelated to the isotropy gauge group, except in a somewhat misleading sense in gravity.)  We can always set $á_{(H)} = d_{(H)}$ in some gauge since $[á_{(H)},á_{(H)}Õ=fá_{(H)}$ (i.e., $F_{(H)(H)}=0$), but in this paper we'll consider the alternative.
  
Our treatment of general curved spaces is a direct generalization of the coset construction for maximally symmetric spaces (flat or de Sitter).  In this first-quantized approach to (super)gravity the derivatives are gauge covariantized with respect to the superPoincar«e group itself:  The Yang-Mills generators (in some matrix representation) are effectively replaced with derivatives with respect to all the coordinates:
$$ d_A £ á_A = e_A{}^M »_M = öe_A{}^M d_M $$
$$ [á_A,á_BÕ = T_{AB}{}^C á_C $$
where $e$ (or $öe$) is now arbitrary.  Acting on a representation of the Lorentz derivatives $á_{(H)}$, the coefficients of $á_{(H)}$ in the covariant derivatives $á_{(G/H)}$ are called ``Lorentz connections"; coefficients of $á_{(H)}$ in the torsion term $Tá$ (i.e., $T_{(G/H)(G/H)}{}^{(H)}$) are called ``curvatures".  Local Lorentz transformations are now included with the rest of the coordinate transformations:
$$ á' = e^ñ áe^{-ñ},ââñ = ñ^M d_M = öñ^A á_A $$
The usual action of local Lorentz transformations on the coset part $á_{(G/H)}$ of the covariant derivative is fixed by the commutator $[á_{(H)},á_{(G/H)}Õ$, whose torsions take their free values (as again does $[á_{(H)},á_{(H)}Õ$).

In this paper we reorganize this procedure:  The isotropy constraints are applied as part of the last step of defining the theory, along with the field equations and other constraints.  We thus have ``covariant" derivatives $d_A$ for the full group ($G$), and they are actually ÓinvariantÕ under the full symmetry group (since left and right multiplication commute, by associativity).  The usual covariant derivatives can be obtained by using the isotropy constraints ($d_{(H)}$) to fix a (unitary) gauge by eliminating the conjugate coordinates.  However, if we do not fix the gauge for the Lorentz constraint, but leave the Lorentz coordinates, the remaining derivatives (for translations and supersymmetry) retain their invariance under Lorentz transformations.  The covariant Lorentz derivatives (``spin") then vanish (for scalars), or equal a particular matrix representation, when acting on a particular superfield.  This is the first-quantized method of defining spin as differential operators, by giving fields (and invariant derivatives) simple, fixed dependence on spin coordinates.  (Similar remarks apply to scale weight in conformal theories derived from higher dimensions [1], which appears in the field as an extra coordinate to the power of the scale weight.)

More importantly, we gauge covariantize the Yang-Mills and supergravity covariant derivatives ÓbeforeÕ applying the isotropy constraints.  Some consequences are: (1)¼the introduction of new gauge fields for derivatives with respect to the coordinates of the isotropy group, which function as prepotentials and (in lightcone gauges) Hertz potentials and compensating gauge parameters; (2) in supergravity, the treatment of the local Lorentz group on a par with general coordinate transformations, so Lorentz generators act as first-quantized operators, and curvatures become additional torsions; and (3) in first quantization, an extension of the affine Lie algebra of the superstring, which also appears for the (gauge-fixed) superparticle off shell, and in the supergravity covariant derivatives as new central charges, making it a direct extension of the algebra of the super Yang-Mills covariant derivatives (see next subsection).

As a generalization, one can define the Lorentz representation by applying not all the Lorentz derivatives as constraints, but some subalgebra that includes a Cartan subalgebra.  
A similar analysis is applied in the corresponding approach to coordinates for R-symmetry (which can be derived from higher-dimensional Lorentz symmetry) in projective [2] and harmonic [3,4] superspaces.  (Similar Lorentz components for Yang-Mills gauge fields have been considered previously [5,6], but with a different analysis of the constraints.  Such coordinates also appear in the pure spinor formalism [7] as the coset space SO(10)/U(5), but without corresponding gauge field components.) 
Although the procedure may look noncovariant, since we pick only a Lorentz subgroup, covariance is preserved simply because each constraint is invariant under the full symmetry group.  Fewer constraints means a larger superspace; noncovariance is avoided because not all Lorentz coordinates can now be gauged away (but dependence on them is fixed by field equations).  The advantage is that some supersymmetry derivatives can be added to the constraints, as long as the constraint algebra still closes.  (If all Lorentz derivatives were constrained, only full representations of the Lorentz derivatives could be used.)  As a result, the increase in the number of bosonic coordinates of the coset space can be traded for a decrease in the number of fermionic coordinates.  Note that, unlike most previous approaches, here the additional coordinates are not added to the usual coset superspace superPoincar«e/Lorentz; instead we use the usual coset construction, with superPoincar«e as the symmetry group, but reduce the isotropy group.

The lightcone analysis of this paper can be interpreted in this fashion:  Some of the results may then be applied to a covariant analysis.  However, the solution of the field equations introduces factors of $1/p^+$, which are nonlocal even when treated covariantly, so such equations shouldn't be considered as kinematic constraints.

ÜDimensional reduction and superstrings

Dimensional reduction is a generalization of the usual coset construction:  It imposes (linear) constraints on the symmetry generators $q$, not the covariant derivatives $d$.  As a result, it does not involve the gauge fields.  If these constraints were not originally central to the symmetry group, they become so, by definition:  These constraints reduce the symmetry to the subgroup that preserves them; however, all covariant derivatives remain invariant.  A familiar example is defining translations in extra dimensions as constraints.  In that case, the covariant derivatives corresponding to translations happen to be just translations themsleves, so all that survives of the corresponding Yang-Mills covariant derivatives is scalars (without derivatives).  And the Lorentz symmetry is reduced to Lorentz for the smaller surviving dimensions plus ``internal" symmetry for those eliminated.  For example, D=26 Yang-Mills dimensionally reduced to D=4 gives Yang-Mills plus 24 scalars, while the analogous coset would yield just 4D Yang-Mills without the scalars.

A less familiar example is the construction for the superstring:  
It's known from the constraint approach to the superstring that the affine Lie algebra for the covariant derivatives includes not only stringy generalizations (currents) $D(§)$ and $P(§)$ of the superparticle's $d$ (for supersymmetry) and $p$ (for translations), but a further operator $¯(§)¾Î'(§)$ (where ``$Ê'Ê$" is the derivative with respect to the worldsheet coordinate $§$) needed for closure of the algebra [8,9].  This $¯$ is thus dual to $D$ in the sense that $D$ is a kind of $Î$ derivative while $¯$ is a kind of $Î$ 1-form (over $§$ space), and consequently also in the sense that they are spinors of opposite chirality.  

Specifically, if we look at the fusion of the superparticle's $Ód,dÕ¾p$ and the bosonic string's $[P,P]¾¶'$, then by applying the Jacobi identity for $DDP$ we find the requirement of an $¯$ such that $[D,P]¾¯$ (times $¶(§)$, which we leave implicit) and $ÓD,¯Õ¾¶'$ (but making $¶'(§)$ explicit):
$$ ÓD,[D,P]Õ ¾ [ÓD,DÕ,P] ¾ [P,P] ¾ ¶' $$
(as well as the $©$-matrix Fierz identity that fixes D=3,4,6,10 classically, for minimal supersymmetry).  Writing collectively
$$ D_\A = (D_{\un Œ},P_{\un a},¯^{\un Œ}) $$
we then have an affine Lie algebra (fixed-``time" current commutation relations) of the form
$$ [D_\A(1),D_\B(2)Õ = -i¶'(1-2)ú_{\A\B} + ¶(1-2) f_{\A\B}{}^\C D_\C $$
where the arguments refer to different values of $§$, and the torsions (structure constants) $f$ are formed from factors of the Minkowski metric and $©$ (generalized Pauli) matrices.  (Underlined Roman indices are Lorentz vector, underlined Greek indices are Lorentz spinor.)  These commutation relations can be solved in terms of the usual coordinates $X$ and $Î$, derivatives $¶/¶X$ and $¶/¶Î$ with respect to them, and their $§$ derivatives $X'$ and $Î'$.  (The $¶'$ includes a $1/Œ'$ for dimensional analysis if the $D$'s are normalized as derivatives, or $Œ'$ if they're normalized as currents.)

Furthermore, in the Lagrangian approach these currents already occur for the (gauge-fixed) superparticle off shell, with $Î'(§)$ replaced with $ÀÏ( )$ (where ``$À{\phantom n}Ê$" is the derivative with respect to the worldline ``time" $ $), as it appears in all of $D,P,¯£d,p,¿$ [10].  
Although $ÀÏ$ might seem trivial, since it vanishes on shell (in an appropriate gauge), its propagator with the conjugate variable $¹$ generates a $¶( )$, which is responsible for 4-point interactions.  In other words, the external field coupling $ÀÏ£ÀÏ+W$ (for spinor field strength $W$) makes its equation of motion nontrivial.

This is also clear from an operator product point of view, since the only differences in operator products for the free (after gauge fixing) operators of the superparticle and string are:  (1) For the superparticle we use the operators $Àx$, $Ï$, $ÀÏ$, and $¹$, functions of only $ $, at unequal times, while for the superstring we use the chiral $»_z X$, $Î$, $»_z Î$, and $¸$, functions of only $z$; and (2) the propagators are different in the two cases, as a result of which the $ú$ term will disappear in the particle case if we restrict the operator products to equal-time commutators, but at general times the algebra will be the same (with different coordinate dependence in the coefficients).

The $DP¯$ affine Lie algebra of the superstring can also be derived as a generalization of an ordinary $dp¿$ Lie algebra for the superparticle defined by the above $f$ by introducing coordinates $Z^\M(§)$ for the group [11]:  We now write
$$ d_\A = (d_{\un Œ},d_{\un a},÷d^{\un Œ}) = (d_{\un Œ},p_{\un a},¿^{\un Œ}) $$
(and similarly for $q_\A$) where $÷d^{\un Œ}$ is ``dual" to $d_{\un Œ}$ with respect to the above group metric $ú_{\A\B}$.  But the coordinates for $÷D^{\un Œ}$ are redundant, and can be eliminated by constraining the ``symmetry generators" $÷Q^{\un Œ}(§)$ to vanish [12] (i.e., dimensional reduction of the fermionic coordinates).  This forces the symmetry generators $Q_{\un Œ}(§)$ to be reduced to just their zero-modes $q_{\un Œ}$ (since $ÓQ_{\un Œ}(1),÷Q^{\un º}(2)Õ¾¶_{\un Œ}{}^{\un º}¶'(1-2)$), the usual supersymmetry.  

A similar construction works for Type II superstrings, introducing separate left and right-handed $DP¯$ algebras and coordinates, with a reduction constraint [13] 
$$ ÷Q_{\un a} ­ Q_{L\un a} - Q_{R\un a} = 0 $$
to identify the bosonic coordinates for both handednesses; the remaining $q_{\un a}$ symmetry is just the zero-modes of $Q_{\un a}­Q_{L\un a}+Q_{R\un a}$, i.e., the usual translations $p_{\un a}$, since $[Q_{\un a},÷Q_{\un b}]¾ú_{\un a\un b}¶'$.  (This can be done for all strings, but for the heterotic case one handedness has only $P$ in its algebra, and Type I can be described as one-handed.)
Thus for the string one generally starts with twice as many coordinates as symmetries, and half the symmetries (the ``dual" ones) are constrained, to ban all but the zero-modes of the other half as symmetries.  This construction can also be applied to the superparticle, and yields a description of supergravity with manifest T-duality [14].  

Note that left and right-handedness on the worldsheet are not the same as left and right multipication of the group elements:  The symmetry currents are always associated with the on-shell modes, while the isotropy currents are associated with modes of the opposite worldsheet handedness.  (This is related to unitarity [12], since for supersymmetry $Óq,qÿÕ=EÊ$ but $Ód,dÿÕ=-E$.  This ``wrong sign" for the supersymmetry covariant derivatives $d_Œ$ is exactly why they are ``eliminated" by second-class constraints.)  Equivalently, one treats the group element $g$ for one worldsheet handedness in the same way as $g^{-1}$ for the other, giving the usual opposite signs in the Wess-Zumino term.
For the rest of this paper we'll usually ignore dualization of translations.

This affine Lie algebra follows from an ordinary Lie algebra (defined by $f$) that is similar in appearance to that for the super Yang-Mills covariant derivatives and spinor field strength 
$$ á_{\un Œ}, á_{\un a}, W^{\un Œ} $$
(again in D=3,4,6,10) except that the latter algebra has also the bosonic field strength $F_{\un a\un b}$.  This is related to the fact that, with these variables for first-quantization, external super Yang-Mills coupling to the superparticle (by Lorentz invariance and dimensional analysis) is described (see next subsection) by the vector connection coupling as usual to $p$, the spinor connection to $¿$, and the spinor field strength $W$ to $d$, but the antisymmetric tensor field strength $F$ can only couple to the (super)spin $s$, the usual relativistic Pauli term.  (Such a term has proven necessary in covariant quantizations of the superparticle/superstring, although previously it appeared only through ghost coordinates [7,15].)

This asymmetry is resolved when coordinates for the Lorentz group are introduced to define the spin:  We then have not only Lorentz ``covariant derivatives" (the spin operators), analogous to the spinor derivatives $d$, but also a dual Lorentz ``current" of the form $g^{-1}Àg$ in terms of the Lorentz group element $g$ (parametrized by the Lorentz coordinates), analogous to $¿$.  (All of $d$, $p$, and $¿$ will also get dependence on the Lorentz coordinates.)  The affine Lie algebra is then extended to
$$ D_\A = (S_{\un a\un b},D_{\un Œ},P_{\un a},¯^{\un Œ},ê^{\un a\un b}) $$
The requirement of $ê$ follows similarly to that of $¯$:  Instead of the supersymmetry algebra for $d,p$, we now consider the superPoincar«e algebra of $s,d,p$; then applying the Jacobi identity for $SPP$, we find the requirement of a $ê$ such that $[P,P]¾ê$ $(+¶')$ and $[S,ê]¾¶'$ $(+ê)$:
$$ [S,[P,P]] ¾ [[S,P],P] ¾ [P,P] ¾ ¶'¼(+ê) $$
Similarly, the $SD¯$ Jacobi implies $ÓD,¯Õ¾ê$ $(+¶')$.  The components of $f$ then appear as in the table of commutators (fixed mostly by dimension and Lorentz symmetry):
$$ \hbox{\vbox{\offinterlineskip
	\halign{ $¼#¼$ & \hbox{\vrule height14pt depth5pt width.4pt} 
		$¼#¼$ & $¼#¼$ & $¼#¼$ & $¼#¼$ & $¼#¼$ \cr
	f & S & D & P & ¯ & ê \cr
	\noalign{\hrule}
	 S & S & D & P & ¯ & ê \cr
	 D & D & P & ¯ & ê & \cr
	 P & P & ¯ & ê && \cr
	 ¯ & ¯ & ê &&& \cr
	 ê & ê &&&& \cr}}} $$
(Note that $¶'$ and $ê$ now always accompany each other.  This is also true when translations are dualized, so $[P,÷P]$ gives both, but $[P,P]$ and $[÷P,÷P]$ neither, so all ``dual" currents commute with each other.)  As for the other dual symmetries, the dual spin coordinates are eliminated by dimensional reduction $÷Q^{\un a\un b}=0$.

Such an extension is natural for anti de Sitter compactifications, where ``translations" require Lorentz for closure (but the structure constants are modified from the above).

ÜCoupling and superparticles

We now consider coupling of external fields:  Without loss of generality we specialize to the superparticle, which has these same currents in its Lagrangian; we begin with super Yang-Mills.  The Lorentz components of the gauge field will then couple to this dual Lorentz current.  Thus, we not only have a spinor connection of opposite chirality to the spinor field strength, but also an antisymmetric tensor gauge field ``dual" to the antisymmetric tensor field strength.  Being of lowest dimension (1 lower than the vector), this gauge field should be interpreted as the ``prepotential".  
Thus the general coupling in the Lagrangian now takes the form
$$ üF^{\un a\un b}s_{\un a\un b} + W^{\un Œ}d_{\un Œ} + A_{\un a}p^{\un a}
	+ A_{\un Œ}¿^{\un Œ} + üA_{\un a\un b}§^{\un a\un b} $$
where in terms of the full symmetry group element $g$ of translations + supersymmetry + Lorentz the currents $J$ are defined as usual by
$$ -ig^{-1}Àg = 
	p^{\un a}T_{\un a} + ¿^{\un Œ}T_{\un Œ} + ü§^{\un a\un b}T_{\un a\un b} $$
for some matrix representation $T$ of the various generators.  (Before including Lorentz coordinates, we had simply $-i¿^{\un Œ}=ÀÏ^{\un Œ}$ and 
$-ip^{\un a}=Àx^{\un a}+iüÀÏ^{\un Œ}©^{\un a}_{\un Œ\un º}Ï^{\un º}$.)  

So we have the currents, progressively increasing in dimension by 1/2, and the fields doing the same,
$$ \li{ d_\A & = 
	(s_{\un a\un b}, d_{\un Œ}, p_{\un a}, ¿^{\un Œ}, §^{\un a\un b}) \cr
	A_\A & = ( A_{\un a\un b}, A_{\un Œ}, A_{\un a}, W^{\un Œ}, F^{\un a\un b}) \cr} $$
(We use graded antihermitian derivatives, so $p®»_x$ without the $-i$, etc.)  The latter can be considered gauge fields for the former in the operator (or Hamiltonian) approach, so these extended covariant derivatives for super Yang-Mills take the form
$$ á_\A = d_\A +iA_\A = 
	(á_{\un a\un b}, á_{\un Œ}, á_{\un a}, ~á^{\un Œ}, ~á^{\un a\un b})  $$
This differs from earlier treatments in that it includes (1) the spin derivative $s_{\un a\un b}$, and (2)¼the dual spin current $§^{\un a\un b}$, which is necessary for an invertible metric $ú_{\A\B}$ (not the non-invertible Cartan metric), but also treats the bosonic field strength $F^{\un a\un b}$ in the same way as the fermionic one $W^{\un Œ}$.

Note that this antisymmetric tensor Yang-Mills gauge field covariantizes the antisymmetric tensor spin ``covariant derivatives":  We now have vector, spinor, and antisymmetric tensor covariant derivatives, all of which transform under the same Yang-Mills gauge transformations.  We have extended the (super)coordinate space, not the gauge group.  

The interaction in the operator approach comes from a generalized d'Alembertian, using the (inverse) metric $ú^{\A\B}$ that relates quantities of opposite dimension with respect to the average.  The higher-dimension currents vanish on shell in the operator approach for the superparticle, so their ``gauge fields" are really field strengths.  (These currents for the superstring have only $§$ derivative terms, so similar remarks apply for their zero-modes.)

As usual, this analysis can be generalized directly to supergravity (or the closed superstring):  In the Lagrangian approach, supergravity couplings are written as the ``square" of super Yang-Mills couplings.  (In the operator approach, vertex operators are also squares for the superstring, where higher-point contributions can be neglected.)  In particular, the supergravity ``prepotential" now carries the same index structure as the Riemann tensor, appearing as 
$\f18 §^{\un a\un b}§^{\un c\un d}h_{\un a\un b,\un c\un d}$, where $h_{\un a\un b,\un c\un d}$ has dimension 2 less than the metric.

The appearance of this prepotential in the covariant derivatives can be understood from the point of view of the first-quantized action as describing parallel transport along the worldline:  A transport term $Çdz^\M á_\M=Çd ÊÀz^\M á_\M$ shows that not only will the prepotential appear in the Lorentz covariant derivative, but as the coefficient of the dual Lorentz current.  Although for the superstring this current survives on shell, and therefore also in a second-quantized approach, its usefulness for the superparticle requires a further generalization.  (This generalization is trivial, although perhaps pedagogical, for super Yang-Mills.)

For supergravity we then have for this enlarged space (including dual coordinates)
$$ [á_\A,á_\BÕ = T_{\A\B}{}^\C á_\C,ââá_\A = E_\A{}^\M d_\M $$
$$ á' = e^ñ áe^{-ñ},ââñ = ñ^\M d_\M = öñ^\A á_\A $$
where $d_\M$ are the flat-space derivatives, but we can also expand over partial derivatives.  
(This is opposite to some approaches in the literature, where all transformations, including Lorentz, supersymmetry, and even translations are treated as Yang-Mills gauge transformations, so all field strengths are ``curvatures"  [16].)  A more useful concept is then to expand the torsion around its vacuum value, i.e.,
$$ T_{\A\B}{}^\C £ f_{\A\B}{}^\C + T_{\A\B}{}^\C $$

ÜFirst-class constraints

For most of the rest of this paper we concentrate on how these prepotentials contribute to solving constraints in a lightcone (or lightcone-like covariant) formalism.  We first derive results for all free theories from group theoretic considerations of first-quantization.  All free massless supersymmetric theories satisfy a simple set of equations of motion, the subset of the free superconformal equations of motion that don't involve S-supersymmetry or conformal boosts [9,17].  (Equivalent equations appear in pure spinor formulations with regard to the $b$ ghost [18].)  These can be expressed in first-quantized language in terms of $p$, $d$, and $s$ (and the scale weight $w$, which is determined by the superspin; we ignore central charges, which vanish on shell, and don't appear in a superconformal derivation).

Because the equations are highly reducible, the higher-(engineering-)dimension equations follow from the lower-dimension ones by hitting with spinor derivatives (the reverse of the procedure of deriving them by applying S-supersymmetry to the Klein-Gordon equation).  It's more convenient to solve them from the top down in a lightcone analysis, where we pick out the irreducible pieces that have been projected by replacement of a free vector index by $+$, and multiplication of spinor equations by the corresponding vector component of the Dirac matrices $©^+$.  (These are really the Pauli matrices, since we use Weyl spinors; we normalize $Ó©^{\un a},©^{\un b}Õ=-ú^{\un a\un b}$.  As usual the lightcone basis is defined by denoting two null directions of a vector by ``$+$" and ``$-$", where we normalize $ú^{+-}=-1$.)  

We divide the constraints into two subsets:  
(1) The first, which we call ``field equations", always have a factor of momentum, and thus can be solved easily in a lightcone analysis by picking out the $p^+$ term:
\vskip-.1in
$$ \li{ p^2 & = 0âÜâp^- = {(p^i)^2\over 2p^+},ââââ¼Êâââgauge¼x^+ = 0 \cr
	Öpd & = 0âÜâ©^- d = {1\over p^+} ©^i p^i ©^- (©^+ d),ââ¼Êgauge¼©^+ Ï = 0 \cr
	s^{\un a \un b}p_{\un b} + wp^{\un a} & = 0âÜâs^{i-} = {1\over p^+}s^{ij}p^j,
		âs^{+-} = w,âgauge¼s^{+i} = 0 \cr} $$
The scale weight $w$ is determined in terms of the spin representation as that which gives a nontrivial solution.  In supersymmetric theories the first solution is redundant to the second, since $Ó©^- d,©^- dÕ=-©^- p^-$ (from $Ód,dÕ=©^{\un a}p_{\un a}$).

The first (Klein-Gordon) equation is the field equation that determines dependence on the lightcone ``time" $x^+$.  The second (``$û$-symmetry") and third determine the dependence on the ``non-lightcone" fermionic coordinates $©^+Ï$ and the longitudinal spin components $s^{ài},s^{+-}$.  The gauge choices are unitary transformations in field theory; for the time this is the usual time-development operator.  (The equations should be imposed on field strengths, but in the lightcone gauge these are the same as the gauge fields up to factors of $p^+$.  The only effect is to shift $w$ to 0 for the gauge field.)  There are also versions of the free equations directly in terms of symmetry generators (which is the form given by applying S-supersymmetry to Klein-Gordon), but the covariant derivative form is that obtained from actions; in particular, the symmetry-covariant derivatives (actually symmetry ÓinvariantÕ before fixing Lorentz coordinates) are gauge covariantized for interactions.  The solutions to the field equations can be used to replace derivatives with respect to non-lightcone coordinates in the symmetry generators, and the corresponding gauge choices to replace the coordinates themselves.

For example, the last equation applied to a vector gives
$$ s^{i-}A^j = {1\over p^+}p^k s^{ik}A^j = {1\over p^+}(p^j A^i - ¶^{ij}p^k A^k) $$
where we have used
$$ s^{ij}A^k = A^{[i}¶^{j]k} ­ A^i ¶^{jk} - A^j ¶^{ik} $$

(2) The second subset, which we call ``representation constraints", further restrict the dependence on the surviving fermionic coordinates $©^-Ï$:
$$ \li{ d ©^{\un a\un b\un c} d + üs^{[\un a\un b}p^{\un c]} & = 0 \cr
	(ü©^{\un a\un b}s_{\un a\un b} + w')d & = 0 \cr} $$
(where $w'$ is another constant, related to the scale weight.)  For cases with nonvanishing spin, the former is redundant to the latter.  They take the place of second-class constraints.  There can be additional terms for R-symmetry:  In this paper we mostly restrict to 10D minimal supersymmetry (super Yang-Mills) to avoid them, but they can be obtained by dimensional reduction.  (In superconformal field theory, such constraints are known as ``semi-shortening" conditions; on the other hand, ``shortening" conditions are the usual isotropy constraints.)

For example, reducing to D=4 and then truncating to N=1, for the case of a scalar with nonvanishing R-symmetry U(1) charge $Y$, the latter simplifies to
$$ (©_{-1}Y + w')d = 0 $$
in Dirac spinor notation, where ``$©_{-1}$" is the product of all the 4D $©$ matrices.  This constraint then states that the scalar is a chiral superfield, and its complex conjugate antichiral.

For the case of 10D super Yang-Mills, after applying the lightcone analysis to the field equations, the latter reduces on a vector to
$$ (ü©^{ij}s^{ij} +w')(©^+ d)A^k = 0âÜâ(©^+ d)A^i - tr = 0 $$
where ``$tr$" is the $©$-matrix trace.  We can also consider its action on a spinor,
$$ (ü©^{ij}s^{ij} +w')(©^+ d)(©^- A) = 0âÜâ(©^+ d)(©^- A) - tr = 0 $$
(again a $©$-trace, i.e., subtracting the one-$©^i$ term), which is related to the vector case by triality.  Solution of these equations requires a second halving of the $Ï$'s:  We'll do this analysis for these cases below.

Û4 Examples

ÜBosonic Yang-Mills

Nontrivial gauge-covariant Lorentz derivatives already exist in Lorentz-noncovari\-ant gauges for nonsupersymmetric theories, although they have generally gone unnoticed.  In a lightcone approach, the nontriviality of the gauge fields for the ``Lorentz derivatives" appears as part of the nonlinearity (in fields) of the Lorentz symmetry generators.  (The remainder of the nonlinearity is due to solving the field equations, as in lightcone formulations of nongauge theories.)  The ``spin" terms in such operators (i.e., terms that differ from their action on color-singlet scalars) include ``compensating gauge transformations" that restore the gauge condition after the Lorentz transformation.  But the spin operators are just the covariant derivatives (the ``isotropy" constraints define the spin representation), so solving the constraints on the covariant derivatives is an equivalent approach to this problem.

For example, consider pure nonsupersymmetric Yang-Mills, with covariant derivatives $á$ for both translations and Lorentz transformations, and field strengths $F$
$$ á = d + iA,ââ[á^{\un a},á^{\un b}] = iF^{\un a\un b},ââ
	[á^{\un a},á^{\un b\un c}] = ú^{\un a[\un b}á^{\un c]} $$
(We will forgo introducing $~á^{\un a\un b}=§^{\un a\un b}+iF^{\un a\un b}$.)  We now use $d$ for all free covariant derivatives (distinguished by indices), and $»$ for all partial derivatives.  (Lorentz derivatives have the implied commutators with themselves, and their field strengths vanish.)  In particular,
$$ [á^{i-}, á^j] = -¶^{ij}á^- âÜâd^{i-}A^j =  -¶^{ij}A^- + á^j A^{i-} $$
The first term is the usual covariant spin term, while the second can be recognized as a gauge transformation.  The full symmetry generator $J^{i-}$ is then the sum of this spin piece and the usual ``orbital" piece $x^{[i}»^{-]}­x^i »^- - x^- »^i$.  ($A^-$ and $»^-$ are then determined by the field equations, as usual.)  Thus in the standard approach $A^{i-}$ appears surreptitiously as these compensating gauge parameters themselves.  In covariant langauge, there is no ``repair" of the Lorentz transformation warranted by change of gauge; the ``gauge-covariant" spin operator automatically preserves the gauge.

Since the other components of the covariant Lorentz derivatives and symmetry generators are linear, their action is the naive one, and we can avoid discussing them, and introducing coordinates for them.  (Alternatively, we can eliminate them by constraints, defining an appropriate coset space [5].  However, rather than just the Lorentz space being a coset, we recognize the whole space as being a coset of Poincar«e.)  The free covariant derivative $d^{i-}$ can then be taken to be just a partial derivative with respect to the remaining coordinates $y^{+i}$ in an appropriate representation.  The commutators
$$ [d^{i-},d^+] = -d^i, ââ[d^{i-},d^j] = -¶^{ij}d^-,ââ[d^{i-},d^-] = 0 $$
then determine the $y$ dependence:  For $»^{i-}y^{j+}=-¶^{ij}$,
$$ \li{ d^{i-} & = »^{i-} \cr
	d^- & = »^- \cr
	d^i¼& = »^i + y^{i+}»^- \cr
	d^+ & = »^+ + y^{i+}»_i + ü(y^{i+})^2»^- \cr} $$
(Similar remarks apply to the spinor derivatives in the supersymmetric case.)  Since dependence on $y$ is completely determined, in the end we'll keep only the $y^{i+}=0$ part of the fields, replacing the free vector covariant derivatives with the usual partial derivatives.  (This is equivalent to gauging $y^{i+}$ away with the constraints that fix the action of $d^{i-}$ on the fields.)

So far we haven't explicitly imposed a gauge.  In covariant gauges the gauge fields for the Lorentz derivatives vanish, so the above discussion is vacuous.  However, in the lightcone gauge
$$ A^+ = 0 $$
we can use Lorentz transformations to relate $A^{i-}$ to $A^i$:
$$ [á^{i-}, á^+] = -á^iâÜâA^i = d^+ A^{i-}âÜâA^{i-} = {1\over d^+} A^i $$
which fixes the explicit form of the compensating gauge parameters.  If we treat instead $A^{i-}$ as fundamental, we can write
$$ d^{i-}A^{j-} = {1\over d^+}( -¶^{ij}A^- + d^{(i}A^{j)-} -i [A^{i-}, d^+ A^{j-}]) $$
This result will find frequent use for supersymmetric generalizations in the following sections.

The free parts of these results agree with the first-quantized result above for $s^{i-}A^j$, after making the identification
$$ A^- = {1\over d^+}d^i A^i = d^i A^{i-} $$
which is the solution to its free field equation.

Another interpretation of these expressions for $A^i$ and $A^-$ in terms of $A^{i-}$ is based on the fact that (in the free case) the lightcone gauge is a special case of the Landau gauge (because of $A^-$'s free field equation).  This is implemented here by the prepotential acting as the (Lorentz covariant) Hertz potential:
$$ d_{\un a}A^{\un a} = 0âÜâA^{\un a} = d_{\un b}A^{\un b\un a} $$
In the lightcone gauge, where $A^{i-}$ is the only survivng part of $A^{\un b\un a}$, we then have
$$ A^+ = d_{\un b}A^{\un b+} = 0,âA^i = d_{\un b}A^{\un bi} = d^+ A^{i-},â
	A^- = d_{\un b}A^{\un b-} = d^i A^{i-}  $$

In the interacting case, the form of the field equation for the ``auxilary" field $A^-$ and the ``physical" fields $A^i$ depends on the choice of interactions.  However, the auxiliary one is generally simpler, and Lorentz invariance can be used to derive the rest from it.  Consider a more general solution of the form
$$ A^- = {1\over d^+}(d^i A^i +ë) $$
where $ë$ is $(1/d^+)J^+$ in terms of the ``current" $J^+$, quadratic in (Yang-Mills and other) fields (in the lightcone gauge, for minimal coupling).  We then solve
$$ [á^{i-},á^-] = 0 $$
using the above solution for $d^{i-}A^{j-}$ (and the commutation relations of the $d$'s) to find
$$ 2d^- A^{i-} = {(d^j)^2\over d^+}A^{i-} +i{1\over d^+}[A^{(i|},d^j A^{|j)-}]
	+{1\over d^+}d^{i-}ë +i\left[A^{i-},{1\over d^+}ë\right] $$
($(ij)­ij+ji$; $|¼|$ leaves alone indices between), where the first term is the usual free one, and $d^{i-}ë$ can be evaluated from the above (and similarly derived results for matter).

In the nonsupersymmetric case $A^{i-}$ then seems redundant to $A^i$.  However, in the maximally supersymmetric case we'll see that the usual lightcone superfield appears in $A^{i-}$; the corresponding component of $A^i$ would require an extra factor of $1/d^+$ (which the lightcone formalism already has in abundance).  This follows from dimensional analysis, as the lightcone kinetic term $(1/g^2)Çd^4ÏÊÄõÄ$ implies $Ä$ is dimensionless.  This is also true in the selfdual case, where $1/d^+$ is not required, and would further break Lorentz symmetry.  Another possible argument is that in a covariant background field gauge the four-point amplitude could be expected to take the form $Çd^{16}ÏÊA^4$ (times momentum integrals), which can give the usual component $F^4$ term only if this $A$ is dimensionless.

ÜSelfdual super Yang-Mills

A similar analysis can be made for selfdual (D=4) Yang-Mills, where a related lightcone gauge appears [19], and whose selfduality implies the field equations already in the nonsupersymmetric case.  The main difference from the non-selfdual bosonic case just considered is that, at least in a lightcone gauge, the prepotential is required in the course of solving the constraints.  Thus it seems natural to introduce it from the beginning, and give an explanation for its origin.

The supersymmetric case can be treated by a simple extension of indices [20]:  In spinor notation, the vector derivative becomes the chiral superspace derivative $á^Œ{}_A$, where one Lorentz SL(2) spinor index becomes an index $A=(a,ÀŒ)$ for GL(N|2) generated by $á^A{}_B$ (including also GL(N) R-symmetry, scale, and antichiral supersymmetry and S-supersymmetry), while the index $Œ$ for the other Lorentz SL(2) $á^{Œº}$ is unchanged.  Compared to the non-selfdual case, we roughly have the substitution
$$ \li{ á_{\un a\un b} &â£âá^{Œº}, á^A{}_B \cr
	á_{\un Œ}, á_{\un a} &â£âá^Œ{}_A \cr
	W^{\un Œ}, F^{\un a\un b} &â£âF_{AB} \cr} $$
The explicit relationship to 4D vector notation for the bosonic parts is $\un a=ŒÀŒ$,
$$ ú^{ŒÀŒ,ºÀº} = C^{Œº}ÐC^{ÀŒÀº},ââ
	á^{ŒÀŒ,ºÀº} = -i(C^{Œº}Ñá^{ÀŒÀº} +ÐC^{ÀŒÀº}á^{Œº}) $$
and the same for $F^{\un a\un b}$ (but for selfduality $F^{μ}=0$), etc.  
($C$ is an antisymmetric symbol, also used to raise indices, with $C^{+-}­ÐC^{À+À-}=i$.)  

The constraints on the covariant derivatives are
$$ [á^Œ{}_A,á^º{}_BÕ = C^{Œº}F_{AB},â¼[á^Œ{}_A,á^B{}_CÕ = ¶_A^B á^Œ{}_C,â¼
	[á^Œ{}_A,á^{º©}] = -iüC^{Œ(º}á^{©)}{}_A $$
From now on we use  $à$ mostly for values of the spinor indices $Œ$.  Then we make the classification
$$ \matrix{ 	\noalign{\vskip-.5\baselineskip}
		&&&& \backb{.93 1 1} \cr
		&&& \backb{.93 1 1} & \back{.85 1 1} \cr
		&& \backb{.93 1 1} & \back{.85 1 1} & \backt{.93 1 1} \cr
		&& \back{.85 1 1} & \backt{.93 1 1} & \cr
		&& \backt{.93 1 1} && \cr
		\noalign{\vskip-4\baselineskip}
		á^{++} && á^{+-}, á^A{}_B && á^{--} \cr
		& á^+{}_A && á^-{}_A & \cr
		&& F_{AB} && \cr
		\noalign{\vskip.5\baselineskip}}ââââââ
	\matrix{ & \backb{.93 1 1} \cr
		& \back{.85 1 1} \cr
		& \backt{.93 1 1} \cr
		\noalign{\vskip-3\baselineskip}
		á_0 & \cr & \O_1 \cr} $$
``Scale weight", which we define to here include R-symmetry (as the GL(1) factor of the GL(N|2)) increases downward, ``Lorentz weight" $s^{+-}$ ($á^{+-}$ for undotted spinor indices $à$, using only 1 of the SL(2)'s) to the right.
It's actually the sum of these two eigenvalues that defines the separation of variables:  We divide these operators up according to whether this eigenvalue sum is 0 (or less; free) or 1 (linear in prepotential).

We then make a similar division for the commutation relations,
$$ \li{ [á_0,á_0Õ & = á_0âÜâlittle¼group\hbox{\it :}¼gauge¼A_0 = 0 \cr
	[á_0,\O_1Õ & = 0â¼¼Üârepresentation \cr
	[á_0,\O_1Õ & = \O_1âÜâ\hbox{\it defines}¼\O_1¼linear¼in¼prepotential \cr
	[\O_1,\O_1Õ & = 0â¼¼Üâ\hbox{\it field}¼equations \cr} $$
to solve order-by-order in the prepotential, in the order listed.  We're really interested only in the equations that involve only covariant derivatives; the field strengths are then defined in the usual way.

The gauge fields $A_0$ in the covariant derivatives $á_0$ appear above and to the left of the basic fields $\O_1$ appearing in the diagonal along the lower right edge.  All of $A_0$ can be gauged to zero, since they generate no nonvanishing field strengths among themselves.

The commutators of $á_0$ with $\O_1$ are then divided up:  First are those that vanish (except for $á_0$'s with vanishing scale weight, which are boring now), which identify the little-group representation of the prepotential, which in this case is $A^{--}$.  Then come those that define the other gauge fields and field strengths $\O_1$ linearly in terms of it:  Since $á^+{}_A$ has vanishing gauge fields, it (actually just $d^+{}_A$) can be used to relate entries in the table by moving in the direction indicated by its position relative to the ``origin" $(á^{+-}, á^A{}_B)$.  Thus it takes us from $á^{--}$ (specifically $A^{--}$) to the other entries in that diagonal one step at a time.  Finally, the commutators of $\O_1$ with itself then give the field equations.  

Beforing considering Lorentz derivatives, the lightcone separation is then
$$ \li{ [á^+{}_A,á^+{}_BÕ & = 0âââÜâA^+{}_A = 0â(little¼group) \cr
 [á^{(+}{}_A,á^{-)}{}_BÕ & = 0âââÜâA^-{}_A = d^+{}_A A^{--}â(representation) \cr
 [á^{[+}{}_A,á^{-]}{}_BÕ & =  2iF_{AB}âÜâF_{AB} = 
 	d^+{}_A d^+{}_B A^{--}â(\hbox{\it definition}) \cr
 [á^-{}_A,á^-{}_BÕ & = 0âââÜâ
 	d^{[-}{}_A d^{+]}{}_B A^{--} +i [d^+{}_A A^{--},d^+{}_B A^{--}Õ = 0 \cr} $$
The representation constraint says the other potentials are ``curl-free", so its solution is in terms of some prepotential $A^{--}$.  The last constraint consists of the field equations that determine the nonlinear dependence on the ``non-lightcone" coordinates, whose derivatives are $d^-{}_A$ (while $d^+{}_A$ remain arbitrary).

Note that the selfdual lightcone gauge condition, instead of breaking both of the manifest Lorentz SL(2,R)'s to GL(1), breaks only one, avoiding the explicit appearance of the usual lightcone's ubiquitous $1/d^{+À+}$.  This and the natural use here of chiral (and not antichiral) superspace make $A^{--}$ the unambiguous prepotential.

However, the existence of this prepotential is already guaranteed without solving differential constraints, as the potential for one of the Lorentz derivatives $á^{Œº}$ [7].  (Such potentials were also considered in [21], but with a different constraint analysis.)  In the above gauge we find, instead of solving the above differential representation constraint, the algebraic definition
$$ [á^{--},á^+{}_A] = -á^-{}_AâÜâA^-{}_A = d^+{}_A A^{--}â(new¼\hbox{\it definition}) $$
Thus we have replaced one nontrivial derivative $á^-{}_A$ with a Lorentz derivative $á^{--}$ in the representation constraint.

Again, as seen from the free part, the free chiral superspace derivatives $d$ in $á=d+A$ are no longer just partial derivatives, but now have some dependence on the Lorentz coordinate $y^{++}$, with $d^{--}=»^{--}$ ($»^{--}y^{++}=-1$),
$$ \li{ d^-{}_A & = »^-{}_A \cr
	d^+{}_A & = »^+{}_A +y^{++}»^-{}_A \cr} $$
 
The field equations can also be replaced, by again substituting $á^{--}$ for one $á^-{}_A$:
$$ [á^{--},á^-{}_A] = 0âÜâ
	d^+{}_A d^{--}A^{--} -2d^-{}_A A^{--} +i[A^{--},d^+{}_A A^{--}] = 0 $$
(We have used the free part of the previous commutator to push the $d^{--}$ to the right of the $d^+{}_A$.)  The solution of either version of the field equations requires breaking the other SL(2):  For the old constraint, we set $B=\rdt+$ (upper; the rest is redundant), and choose $A=a$ or $À-$ to solve for the non-lightcone-$Ï$ derivatives $»^-{}_a$ and for the time development $»^{-À-}$.  For the new constraint we now instead first set $A=\rdt+$ (upper) to solve for $d^{--}$, then plug this solution into the equation for the other values of $A$ for the same as before.  The results are
$$ \li{ d^{--}A^{--} & = {1\over d^{+À+}}(2d^{-À+}A^{--}+i[d^{+À+}A^{--},A^{--}]) \cr
 d^-{}_a A^{--} & = 
 	{1\over d^{+À+}}(d^{-À+}d^+{}_a A^{--} +i[d^{+À+}A^{--},d^+{}_a A^{--}Õ) \cr
 d^{-À-}A^{--} & = 
 	{1\over d^{+À+}}(d^{-À+}d^{+À-}A^{--} +i[d^{+À+}A^{--},d^{+À-}A^{--}Õ) \cr} $$
The first is identical to the nonsupersymmetric case, as a component of what we found for $d^{i-}A^{j-}$, since it doesn't involve spinors.    The last is a rearrangement of $õA^{--}=...$ .

The generalization to the non-selfdual (4 $²$ D $²$ 10) case is similar (but more complicated: see the next section).  In particular, the non-selfdual action in D=4 is most conveniently written as the selfdual one plus extra terms, especially for an expansion in order of helicity violation [22].  (Extension to D>4 then adds yet further terms.)  This implies the appropriateness of the selfdual prepotential for the general case.

ÜSelfdual supergravity

A similar construction can be used for selfdual supergravity [20].  The one we describe here applies to the (fully) gauged version.  (The vectors have Yang-Mills gauge group SO(N) for N supersymmetries.)  Because of the gauging, the GL(N|2) of the selfdual Yang-Mills case is replaced with its subgroup OSp(N|2).  (The result is the chiral contraction of OSp(N|4), namely I[OSp(N|2)$°$Sp(2)], with no cosmological constant because of the selfduality.)  We can now make use of the central charges similar to those of the non-selfdual case described above; unlike the Yang-Mills case (where we ignored them), they will play an important role here.  The relation to the non-selfdual flat derivatives is roughly
$$ \li{ s_{\un a\un b} &â£âs^{Œº}, s^{AB} \cr
	d_{\un Œ}, p_{\un a} &â£âd^Œ{}_A \cr
	¿^{\un Œ}, §^{\un a\un b} &â£â§_{Œº}, §_{AB} \cr} $$
where $s^{AB}$ are the graded-antisymmetric OSp(N|2) generators, related to the GL(N|2) generators $s^A{}_B$ as  
$$ s^{AB} ­ s^{[A}{}_C ú^{B)C} $$
Note that this reduced symmetry was already suggested in the Yang-Mills case since $s^{AB}$, which carries the same symmetry on its indices as $§_{AB}$, now in turn carries the same as the field strength $F_{AB}$.  Also, now the indices can be freely raised and lowered by the graded-symmetric OSp metric $ú^{AB}$ and its inverse $ú_{AB}$ (and of course we can also use the metric $C$ for the other SL(2)).  

The structure constants of the Lie algebra correspond to the table
$$ \hbox{\vbox{\offinterlineskip
	\halign{ $¼#¼$ & \hbox{\vrule height14pt depth5pt width.4pt} 
		$¼#¼$ & $¼#¼$ & $¼#¼$ \cr
	f & s & d & § \cr
	\noalign{\hrule}
	 s & s & d & § \cr
	 d & d & § & \cr
	 § & § && \cr}}} $$
(I.e., $s$ acts as OSp(N|2)$°$Sp(2) on everything according to their indices, including itself, but also $[d,dÕ=§$.)  We thus consider the derivatives
$$ \matrix{ 	\noalign{\vskip-.5\baselineskip}
		&&&& \backb{.93 1 1} \cr
		&&& \backb{.93 1 1} & \back{.85 1 1} \cr
		&& \backb{.93 1 1} & \back{.85 1 1} & \backt{.93 1 1} \cr
		&& \back{.85 1 1} & \backt{.93 1 1} & \cr
		&& \backt{.93 1 1} && \cr
		\noalign{\vskip-4\baselineskip}
		á^{++} && á^{+-}, á^{AB} && á^{--} \cr
		& á^+{}_A && á^-{}_A & \cr
		~á^{++} && ~á^{+-},~á_{AB} && ~á^{--} \cr
		\noalign{\vskip.5\baselineskip}}ââââââ
	\matrix{ \noalign{\vskip-\baselineskip}
		&& \backb{.93 1 1} \cr
		& \backb{.93 1 1} & \back{.85 1 1} \cr
		\backb{.93 1 1} & \back{.85 1 1} & \backt{.93 1 1} \cr
		\back{.85 1 1} & \backt{.93 1 1} & \cr
		\backt{.93 1 1} && \cr
		\noalign{\vskip-4\baselineskip}
		á_0 && \cr & \O_1 & \cr && \O_2 \cr} $$

\noindent where now there is the introduction of $\O_2$, not appearing in the selfdual super Yang-Mills case.  

The nontrivial (i.e., not involving $s$) commutators are now
$$ [á^Œ{}_A,á^º{}_BÕ = ú_{AB}~á^{Œº} +C^{Œº}F_{AB}$$
 $$ iF_{AB} ­ ~á_{AB} +üR_{ABCD}á^{DC} $$
the nontrivial part being the $R$ term.  A sufficient set of constraints is
$$ \li{ [á^+{}_A,á^+{}_BÕ & = 0âââÊÜâá^+{}_A = d^+{}_A \cr
 [á^{--},á^+{}_A] & = -á^-{}_A¼¼ÊÜâá^-{}_A = [d^+{}_A, á^{--}] \cr
 [á^{[+}{}_A,á^{-]}{}_BÕ & =  2iF_{AB}âÜâ
 	representation¼and¼\hbox{\it definition} \cr
 [á^{--},á^-{}_A] & = 0âââÊÜâ\hbox{\it field}¼equations \cr} $$
These are similar to the selfdual Yang-Mills case, except for the restriction on the form of $F_{AB}$.  ($R$ could be absorbed by a redefinition of $~á$, but then would pop up in other commutators.)  Any new ones involving $\O_2$ are redundant or definitions.
Since we won't consider non-selfdual supergravity in much detail, we'll concentrate on using the solution of the first two equations to solve the third.  

The solution is essentially the same as that given previously [19], with again the difference that the prepotential already appears in the covariant derivatives, rather than being discovered by solving a differential equation as in the original treatment of the bosonic case [23].  The result is
$$ \li{ á^{--} & = s^{--} +ü(d^{+A}d^{+B}h^{-4})s_{BA} +2(d^{+A}h^{-4})d^+{}_A
		+3h^{-4}§^{++} \cr
	á^-{}_A & = d^-{}_A +ü(d^+{}_A d^{+B}d^{+C}h^{-4})s_{CB} 
		+(d^+{}_A d^{+B}h^{-4})d^+{}_B +(d^+{}_A h^{-4})§^{++}\cr
	iF_{AB} & = §_{AB} 
		+ü(d^+{}_A d^+{}_B d^{+C}d^{+D}h^{-4})s_{DC} \cr
	& ÜâR_{ABCD} = d^+{}_A d^+{}_B d^+{}_C d^+{}_D h^{-4} \cr} $$
where we have abbreviated $h^{-4}$ for $h^{----}$.  The relative coefficients of the $h$ terms in $á^{--}$ are fixed so that with each commutator with $d^+{}_A$ there is some cancellation between a term with it acting on $h$ and a term with its commutator with a derivative:  Multiple commutators with $d^+{}_A$ act as
$$ d^+{}_A:¼\leftÓ \matrix{ s^{--} & £ & d^-{}_A & £ & §_{AB}â(§^{+-}) \cr
				s_{AB} & £ & d^+{}_A & £ & §^{++}\hfill \cr} \right. $$

Note that, whereas in the Yang-Mills case the prepotential appears only as its derivative if the Lorentz derivative isn't introduced, in the supergravity case it appears only as its ÓsecondÕ derivative if both the Lorentz derivative and the dual Lorentz central charge aren't introduced:  $h^{-4}$ appears without derivatives only as the $§^{++}$ term of $á^{--}$.

ÜN=2 and general

The prepotential appears in a similar way [24] in harmonic supergravity [4,25], now also simplifed by the extra currents/coordinates:  We now have the derivatives
$$ \matrix{ á^{++} && á^{+-}, á_Œ{}^º && á^{--} \cr
		& á^+{}_Œ && á^-{}_Œ & \cr
		&& á^{Œº} && \cr
		& ~á^{+Œ} && ~á^{-Œ} & \cr
		~á^{++} && ~á^{+-}, ~á_Œ{}^º && ~á^{--} \cr} $$

\noindent where now $Œ$ is a 6-dimensional spinor index, $á^{Œº}$ is the (6-)vector derivative, and $à$ is for R-symmetry SU(2), as might arise from Lorentz in still higher dimensions.  (For convenience we look at 4D N=2 supersymmetry as 6D N=1, or at least use that notation.)  In the harmonic formalism $á^{++}$ is not gauged to the free value, but still satisfies the ``analyticity" condition
$$ [d^+{}_Œ,á^{++}] = 0 $$
where $á^+{}_Œ$ has been gauged to its free value.  In the Yang-Mills case, this simply states the analyticity of the prepotential $A^{++}$.  In the supergravity case, the solution is now
$$ \li{ á^{++} & = s^{++} +\f1{4!}·^{Œº©¶}Ód^+{}_Œ,[d^+{}_º,Ód^+{}_©,[d^+{}_¶,Us^{--}]Õ]Õ \cr
	& = s^{++} + [(d^+)^4 U]s^{--} + [(d^+)^{3Œ}U]d^-{}_Œ +
	ü(d^+{}_Œ d^+{}_º U)d^{Œº} + (d^+{}_Œ U)¿^{+Œ} + U§^{++} \cr} $$
in terms of the prepotential $U$.  In this case multiple commutators with $d^+{}_Œ$ act as
$$ d^+{}_Œ:âs^{--}¼£¼d^-{}_Œ¼£¼d^{Œº}¼£¼¿^{+Œ}¼£¼§^{++} $$
or more explicitly,
$$ [d^+{}_Œ,s^{--}] = d^-{}_Œ,âÓd^+{}_Œ,d^-{}_ºÕ = d_{Œº},â
	[d^+{}_Œ,d^{º©}] = ¿^{+[º}¶_Œ^{©]},âÓd^+{}_Œ,¿^{+º}Õ = ¶_Œ^º §^{++} $$
$$ (d^+)^4 ­ \f1{4!}·^{Œº©¶}d^+{}_Œ d^+{}_º d^+{}_© d^+{}_¶,â
	(d^+)^{3Œ} ­ \f1{3!}·^{º©¶Œ}d^+{}_º d^+{}_© d^+{}_¶,â
	d^{Œº} ­ ü·^{©¶Œº}d_{©¶} $$
The former arose from the SU(2)-covariant relations (with R-index $a$ on $d^a{}_Œ$)
$$ [d^a{}_Œ,s^{bc}] = -iüC^{a(b}d^{c)}{}_Œ,âÓd^a{}_Œ,d^b{}_ºÕ = -iC^{ab}d_{Œº} $$
$$ [d^a{}_Œ,d^{º©}] = ¿^{a[º}¶_Œ^{©]},âÓd^a{}_Œ,¿^{bº}Õ = ¶_Œ^º §^{ab} $$
Even with the usual harmonic coordinates the prepotential would not appear explicitly with fewer than 2 derivatives without $¿$ and $§$.

As a matter of dimensional analysis, we note that even if the kinetic term of a supersymmetric theory involves only Lorentz (or R) derivatives, which are dimensionless, and the $Ï$ integration is over only half the maximum (as for chiral, projective, or harmonic Lagrangians), the prepotentials in D=4 still must have a dimension (following from $Çd^4 xÊd^{2N}ϼV^2$) of (at most) 2$-$N/2.  For the dimensions of the maximally supersymmetric cases, we then find: 
\item{(a)} 1 for the N=2 scalar multipet, the dimension of a physical scalar; 
\item{(b)} 0 for N=4 Yang-Mills, the dimension of the Lorentz (pre)potential; and 
\item{(c)} $-$2 for N=8 supergravity, the dimension of the dual-Lorentz ($§$) connection appearing in the Lorentz ($s$) derivative.  

\noindent These are also the dimensions and fields found above for N=2 and for general seldual theories.  

On the other hand, in background field gauges we can choose a gauge for the background where these lower-dimensional potentials are set to vanish (since their field strengths vanish, and we don't need to solve constraints for the background), resuting in the usual nonrenormalization theorems [26].

Û4 10D super Yang-Mills

ÜSeparation of equations

In nonsupersymmetric theories, lightcone formulations of gauge theories (both Yang-Mills [27] and gravity [28]) were originally derived by starting with the covariant (under both Lorentz and gauge transformations) formulation, fixing a null gauge, and eliminating nonpropagating modes by their field equations.  On the other hand, in the supersymmetric case the result has been obtained by either (1) combining component results into lightcone superfields [29] or (2) developing free lightcone superspace by symmetry arguments, and preserving consistency upon introducing interactions [30].  In the case of supergravity the latter method has succeeded only to lowest orders in the coupling [31].  This suggests that the covariant (also under supersymmetry) method should be more straightforward.

Here we apply this method to maximally supersymmetric Yang-Mills in ten dimensions.  (Results in lower dimensions, including cases with less supersymmetry, can be derived by the same method or dimensional reduction.)  Extension to supergravity is straightforward (but more tedious, as seen from the nonsupersymmetric case).  Although totally (manifestly) covariant formulations of maximally supersymmetric gauge theories are known only on shell [32,33], in a lightcone analysis it's easy to separate the field equations from the representation constraints.

Our gauge-covariant objects for super Yang-Mills include the spinor ($W$) and antisymmetric tensor ($F$) field strengths (which define the usual component field strengths at $Ï=0$), the usual spinor and vector covariant derivatives (for $Ï$ and $x$), and the antisymmetric tensor covariant derivatives for Lorentz spin introduced above.  In a lightcone analysis we can classify them by both their scale weight and their eigenvalue under the component of the Lorentz spin operator $s^{+-}$:
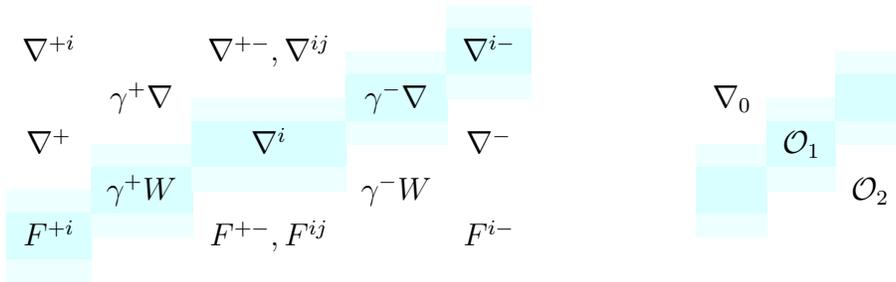
$$ \matrix{ 	\noalign{\vskip-.5\baselineskip}
		&&&& \backb{.93 1 1} \cr
		&&& \backb{.93 1 1} & \back{.85 1 1} \cr
		&& \backb{.93 1 1} & \back{.85 1 1} & \backt{.93 1 1} \cr
		& \backb{.93 1 1} & \back{.85 1 1} & \backt{.93 1 1} & \cr
		\backb{.93 1 1} & \back{.85 1 1} & \backt{.93 1 1} && \cr
		\back{.85 1 1} & \backt{.93 1 1} &&& \cr
		\backt{.93 1 1} &&&& \cr
		\noalign{\vskip-6\baselineskip}
		á^{+i} && á^{+-}, á^{ij} && á^{i-} \cr
		& ©^+ á && ©^- á & \cr
		á^+ && á^i && á^- \cr
		& ©^+ W && ©^- W & \cr
		F^{+i} && F^{+-}, F^{ij} && F^{i-} \cr
		\noalign{\vskip.5\baselineskip}}ââââââ
	\matrix{ \noalign{\vskip-\baselineskip}
		&& \backb{.93 1 1} \cr
		& \backb{.93 1 1} & \back{.85 1 1} \cr
		\backb{.93 1 1} & \back{.85 1 1} & \backt{.93 1 1} \cr
		\back{.85 1 1} & \backt{.93 1 1} & \cr
		\backt{.93 1 1} && \cr
		\noalign{\vskip-4\baselineskip}
		á_0 && \cr & \O_1 & \cr && \O_2 \cr} $$
Scale weight increases downward (Lorentz, spinor, and vector covariant derivatives; and spinor and antisymmetric tensor field strengths), $s^{+-}$ ($á^{+-}$) weight to the right.

As in the (4D) selfdual case, we use the sum of these two eigenvalues to define the separation of variables:  It corresponds to the polynomial order in the prepotential.  We now label these operators for eigenvalue sum 0 (or less; free), 1 (linear in prepotential), or 2 (or higher; nonlinear).  We also classify the commutation relations as before:
$$ \li{ [á_0,á_0Õ & = á_0âÜâlittle¼group\hbox{\it :}¼gauge¼A_0 = 0 \cr
	[á_0,\O_1Õ & = 0â¼¼Üârepresentation \cr
	[á_0,\O_1Õ & = \O_1âÜâ\hbox{\it defines}¼\O_1¼linear¼in¼prepotential \cr
	[\O_1,\O_1Õ & = 0â¼¼Üâ\hbox{\it field}¼equations \cr} $$

The derivatives $á_0$ now appear above and to the left of the main diagonal in the table, which is drawn from the lower left corner to the upper right.  These derivatives correspond to the ``little group" [34] of nonvanishing free derivatives $d_0$ in the lightcone frame $d^i=0$:  There $d^-$, $©^-d$, and $d^{i-}$ also vanish by the (free) field equations, as we saw in the first-quantized treatment.

The basic fields $\O_1$ appear along this diagonal.  The prepotential in this case is a Lorentz component of $A^{i-}$.  Now $©^+d$ and $d^+$ can be used to relate entries in the table (relative to the ``origin" $(á^{+-}, á^{ij})$), taking us from $A^{i-}$ to all the other entries in that diagonal, either one step at a time (by $©^+d$) or two (by $d^+$). 

Just as for $[á_0,\O_1Õ$, the commutators of $\O_1$ with itself are also divided into those that vanish, which give the field equations, and those that define the gauge field and field strengths $\O_2$ to the right of and below the diagonal.  

For the maximally supersymmetric case it's sufficient (but not necessarily preferable) to consider just the spinor-spinor anticommutators $Óá,áÕ = -i©Éá$ (for D=10, or in D=10 notation [32]), as is usually done when Lorentz derivatives aren't considered:
$$ \li{ Ó©^+á,©^+áÕ & = i©^+ á^+ââ¼¼Üâ©^+ A = A^+ = 0 \cr
	Ó©^+á,©^-áÕ & = i©^+©^-©^i á^iâÜâ(©^+ d)(©^- A) = i©^+©^-©^i A^i \cr
	Ó©^-á,©^-áÕ & = i©^- á^-ââ¼¼Üâ\hbox{\it field}¼equations \cr} $$
The other commutators are redundant.  The first anticommutator is the little-group constraint for both $©^+á$ and (by Jacobi identity) $á^+$.  The second anticommutator is a representation constraint plus a definition, which says that (the lightcone part of) the spinor derivative acts as a Dirac matrix (up to factors involving $d^+$) on the two ``Weyl spinors" $©^-A$ and $A^i$ (up to a triality, so that $©^+d$ is the ``vector").  Since it's linear, the representation part agrees with the free first-quantized result.  The last anticommutator gives field equations defining $d^-$ and the nonlinear supersymmetry transformations $©^-q$.

Introducing the covariant Lorentz derivative, we can replace $©^-á$ with $á^{i-}$ in the above to get a different set of sufficient equations.  (The previous then follow via the Jacobi identities.)  This is the same type of replacement of representation constraint made above in the selfdual case (there replacing $á^-{}_A$ with $á^{--}$).  We then look at
$$ [á^{i-},©^+á] = ©^i ©^+(©^-á)âÜâ(©^+d)A^{i-} = ©^i©^+(©^-A)â(new¼representation) $$
which is effectively the same representation constraint, using the anticommutation relations of the ``Dirac matrices" $©^+d$ (again the same as the free first-quantized result, and related to the constraint on the spinor by triality).  We can therefore replace the previous representation constraint with this new, lower-dimension one.

The equivalence is a little clearer if we note that, as in the nonsupersymmetric case above,
$$ [á^{i-},á^+] = -á^iâÜâA^i = d^+ A^{i-} $$
(Similar equations hold for the other entries on the main diagonal of the table.)

We can also make a similar replacement of the field equation constraint, trading one or both of $©^-á$ for $á^{i-}$:
$$ [á^{i-},©^-á] = 0âorâ[á^{i-},á^{j-}] = 0â(new¼\hbox{\it field}¼equations) $$

ÜEuphoric representation

The (4D) selfdual theory can be defined on chiral superspace, since antichiral supersymmetry transformations are so trivial there.  (They are included in the manifest GL(N|2) symmetry of the indices.)  In the more general theory chirality arises from the representation constraint. 

As usual the representation constraint for 10D lightcone super Yang-Mills can be solved by breaking the lightcone SO(8) symmetry to U(4) = SU(4)$°$U(1), corresponding to the standard method of expressing Dirac matrices in terms of creation and annihilation operators.  The SU(4) = SO(6) symmetry is the same as in the 4D N=4 case, except we now keep the derivatives $d_{ab}$ for the 6 extra dimensions.  The U(1) = SO(2) is part of 4D Lorentz:  We can take the 4 ``broken" dimensions (10$-$6 = 4) as those of D=4.  Thus SO(9,1) is broken to SO(6)$°$SO(3,1), and the SO(3,1) is in turn broken to SO(2)$°$SO(1,1), where the SO(1,1) = GL(1) is the $-s^{+-}$ (in vector notation) used above.  

Thus it's convenient to use 4D spinor notation for the 10D theory, so as to indictae explicitly both the SO(1,1) and SO(2) charges, as well as the SU(4).  We therefore convert the $à$ vector notation of the previous subsection to the $à,\rdtà$ spinor notation of the previous section:
$$ \li{ á^+ &â£âá^{+À+} \cr
	á^- &â£âá^{-À-} \cr
	á^i &â£âá^{-À+},¼á^{+À-},¼á_{ab} = ü ·_{abcd}Ñá^{cd} \cr
	á^{i-} &â£âá^{--},¼Ñá^{À-À-},¼á_{ab}{}^{-À-} = ü ·_{abcd}Ñá^{cd-À-} \cr
	©^+á &â£âá^+{}_a,¼Ñá^{À+a} \cr
	©^-á &â£âá^-{}_a,¼Ñá^{À-a} \cr} $$
and the same for the potentials $A$.  The $à$ indices thus still indicate SO(1,1) weight (but now in half steps), while the SO(2) weights are $+ü$ for $+$ and $\rdt-$, and $-ü$ for $-$ and $\rdt+$ (all for upper indices).  Some normalizations we use follow from those of the previous section (and complex conjugation); we use
$$ Ód^Œ{}_a,Ðd^{Àºb}Õ = -i¶_a^b d^{ŒÀº},ââ
	Ód^Œ{}_a,d^º{}_bÕ = -iC^{Œº}d_{ab},ââÓÐd^{ÀŒa},Ðd^{Àºb}Õ = -iÐC^{ÀŒÀº}ü·^{abcd}d_{cd} $$
The latter follow from identifiying $A_{ab}$ with the selfdual scalars $F_{ab}$ after dimensional reduction.  The implied convention for the metric is then, for general 8-vectors $\A$ and $\B$,
$$ \A^i\B^j ¶_{ij} = \A^{+À-}\B^{-À+} + \A^{-À+}\B^{+À-} -\f14 ·^{abcd}\A_{ab}\B_{cd},ââ
	\A_{ab} = -ü·_{abcd}(\A*)^{cd} $$

The representation constraints then become the usual oscillator relations (up to factors of $d^{+À+}$, since $Ód^+{}_a,Ðd^{À+b}Õ=-i¶_a^b d^{+À+}$)):  The chirality and ``reality" conditions are
$$ Ðd^{À+a}A^{--} = 0,ââd^+{}_a d^+{}_b A^{--} = ü·_{abcd}Ðd^{À+c}Ðd^{À+d}ÐA^{À-À-} $$
while the expansion in $Ï^{-a}$ is given by
$$ \li{ \openup1\jot A^{--} & \cr
	d^+{}_a A^{--} & = A^-{}_a \cr
	d^+{}_a d^+{}_b A^{--} & = d^{+À+} A_{ab}{}^{-À-} ââÊ= A_{ab} \cr
	id^+{}_a d^+{}_b d^+{}_c A^{--} & = ·_{abcd}d^{+À+} ÐA^{À-d} â¼= ·_{abcd}W^{+d} \cr
	d^+{}_a d^+{}_b d^+{}_c d^+{}_d A^{--} & = ·_{abcd}(d^{+À+})^2 ÐA^{À-À-} =
		 ·_{abcd}F^{++} \cr} $$
$A^{--}$ can again be used as the chiral-superspace prepotential, as for the selfdual sector, but now we also have the antichiral $ÐA^{À-À-}$.  In the selfdual case, because of Wick rotation, it's independent and vanishing, while $A^{--}$ is real.  (This is what extends the R symmetry from SL(N) to GL(N) in the selfdual case; the extra GL(1) is the continuous duality transformation.)  In Minkowski space for the general case it's the complex conjugate of $A^{--}$ ($ÐA^{À-À-}=(A^{--})*$), while the reality is that of $d^+{}_a d^+{}_b A^{--}$.

Remembering also the relations $A^i=d^+A^{i-}$, $©^+W=d^+©^+©^-A$, and $F^{+i}=d^{+2}A^{i-}$, we then can write all the potentials/field strengths in $\O_1$ in terms of the prepotential as (with normalizations from previous sections)
$$ A_{ab}{}^{-À-} = {1\over d^{+À+}}Êd^+{}_a d^+{}_b A^{--},ââ
	ÐA^{À-À-} = {1\over (d^{+À+})^2}(d^+)^4 A^{--} $$
$$ A^-{}_a = d^+{}_a A^{--},ââÐA^{À-a} =Ðd^{À+a}ÐA^{À-À-} 
	= -i{1\over d^{+À+}}Ê(d^+)^{3a}A^{--} $$
$$ A^{-À+} = d^{+À+}A^{--},ââA_{ab} = d^+{}_a d^+{}_b A^{--},ââ
	A^{+À-} = {1\over d^{+À+}}Ê(d^+)^4 A^{--} $$
$$ ÑW{}^{À+}{}_a = d^{+À+}d^+{}_a A^{--},ââW^{+a} = -i(d^+)^{3a}A^{--} $$
$$ ÐF^{À+À+} = (d^{+À+})^2 A^{--},ââF^{+À+}{}_{ab} = d^{+À+}d^+{}_a d^+{}_b A^{--},ââ
	F^{++} = (d^+)^4 A^{--} $$
where
$$ (d^+)^4 ­ \f1{24}·^{abcd}d^+{}_a d^+{}_b d^+{}_c d^+{}_d,âââ
	(d^+)^{3a} ­ \f16 ·^{abcd}d^+{}_b d^+{}_c d^+{}_d $$

Thus these quantities are related by the action of $d^+{}_a$, $Ðd{}^{À+a}$, and $d^{+À+}$ as indicated by the following table, where scale weight and Lorentz SO(1,1) weight increase downward and Lorentz U(1) to the right:
$$ \matrix{ A^{--} && A_{ab}{}^{-À-} && ÐA^{À-À-} \cr
	& A^-{}_a && ÐA^{À-a} & \cr
	A^{-À+} && A_{ab} && A^{+À-} \cr
	& ÑW{}^{À+}{}_a && W^{+a} \cr
	ÐF^{À+À+} && F^{+À+}{}_{ab} && F^{++} \cr}ââââ
\matrix{ & \swarrowââ\searrow & \cr
	Ðd^{À+a} \hskip-1em && \hskip-1em d^+{}_a \cr
	\noalign{\vskip-1.6\baselineskip}
	& \Bigg\downarrow & \cr
	& d^{+À+} & \cr} $$

ÜField equations

The procedure for solving the field equations is a generalization of that used for the selfdual theory.  (We have already solved the little group and representation constraints.  The linear parts of the following all agree with the first-quantized results above.)  The basic idea is to solve for the action of $d^{i-}$, $©^- d$, and $d^-$ on the prepotential, by looking at commutators of covariant derivatives.  Since the action of $d^i$ is unconstrained, it's more convenient to use $á^{-À+}$ in place of $á^{--}$ in field equations where possible, i.e., when it yields vanishing commutator.  (Remember $A^{-À+}=d^{+À+}A^{--}$.  Another alternative would involve introducing $\O_2$ field strengths, which would be eliminated using their field equations, as derived by higher-order commutators.)

We begin with equations of the form $[á^{i-},á^j]$, which were already solved in the nonsupersymmetric case to find $d^{i-}A^{j-}$:
$$ \li{ [á^{--},á^{-À+}] = 0& ¼¼Üâ
	d^{--}A^{--} = {1\over d^{+À+}}( 2d^{-À+}A^{--} -i[A^{--},A^{-À+}] ) \cr
	[á_{ab}{}^{-À-},á^{-À+}] = 0 & ¼¼Ü¼¼ d_{ab}{}^{-À-}A^{--} = 
	{1\over d^{+À+}}( d^{-À+}A_{ab}{}^{-À-} +d_{ab}A^{--} -i[A_{ab}{}^{-À-},A^{-À+}] ) \cr
	} $$
(We have left $A$'s not explicitly expressed in terms of the prepotential, for conciseness; the representation solution above should be applied.)  
We also have the complex conjugate solutions
$$ \li{ Ðd^{À-À-}ÐA^{À-À-} & = {1\over d^{+À+}}( 2d^{+À-}ÐA^{À-À-} -i[ÐA^{À-À-},A^{+À-}] ) \cr
	d_{ab}{}^{-À-}ÐA^{À-À-} & = 
	{1\over d^{+À+}}( d^{+À-}A_{ab}{}^{-À-} +d_{ab}ÐA^{À-À-} -i[A_{ab}{}^{-À-},A^{+À-}] ) \cr} $$

In the selfdual case we also solved two equations which now have identical solutions: the former of the two above, and 
$$ [á^-{}_a,á^{-À+}] = 0âÜâd^-{}_a A^{--} = {1\over d^{+À+}}( d^{-À+}A^-{}_a 
		-i[A^-{}_a,A^{-À+}] ) $$
The complex conjugate is
$$ Ðd^{À-a}ÐA^{À-À-} = {1\over d^{+À+}}( d^{+À-}ÐA^{À-a}  -i[ÐA^{À-a},A^{+À-}] ) $$

The easiest (and most useful) way to solve for $Ðd^{À-À-}A^{--}$ and $Ðd^{À-a}A^{--}$, rather than solving more constraints, is to write $A^{--}$ in terms of $ÐA^{À-À-}$ using the complex conjugate of the equation for the reverse 
$$ A^{--} = {1\over (d^{+À+})^2}(Ðd^{À+})^4 ÐA^{À-À-},ââ
	(Ðd^{À+})^4 ­ \f1{24}·_{abcd}Ðd^{À+a} Ðd^{À+b} Ðd^{À+c} Ðd^{À+d} $$
and pushing the $Ðd^{À-À-}$ or $Ðd^{À-a}$ to the right till it hits the $ÐA^{À-À-}$.  This requires the special cases of the above commutators,
$$ [Ðd^{À-À-},d^{+À+}] = - d^{+À-},ââ[Ðd^{À-À-},Ðd^{À+a}] = - Ðd^{À-a},ââ
	ÓÐd^{À-a},Ðd^{À+b}Õ = -ü·^{abcd}d_{cd} $$
Using the above solutions for the $Ðd^{À-À-}$'s or $Ðd^{À-a}$'s that reach the $ÐA^{À-À-}$, we find
$$ \li{ Ðd^{À-À-}A^{--} =¼& {1\over d^{+À+}}\f14 ·^{abcd}d_{ab}A_{cd}{}^{-À-} \cr
& -i{1\over (d^{+À+})^3} (Ðd^{À+})^4 [ÐA^{À-À-},A^{+À-}]
	+{1\over (d^{+À+})^4}d^+{}_a (Ðd^{À+})^4 [ÐA^{À-a},A^{+À-}] \cr
Ðd^{À-a}A^{--} =¼& -i{1\over d^{+À+}}ü·^{abcd}d_{bc}A^-{}_d
	-i{1\over (d^{+À+})^3} (Ðd^{À+})^4 [ÐA^{À-a},A^{+À-}] \cr} $$
These two are also simply related by $Ðd^{À+a}Ðd^{À-À-}A^{--}=Ðd^{À-a}A^{--}$.  (We have also used $[d^+{}_a,(Ðd^{À+})^4]=-id^{+À+}(Ðd^{À+})^3{}_a$ to replace a $Ðd^3$ with a $Ðd^4$.)

The complex conjugates of these involve $(d^+)^4$ rather than $(Ðd^{À+})^4$.  We find the latter more convenient, so we instead use
$$ \li{ [ Ñá^{À-À-}, á^{--}] =¼& 0¼Ü \cr
d^{--}ÐA^{À-À-} =¼ 
	& {1\over d^{+À+}}\f14 ·^{abcd}d_{ab}A_{cd}{}^{-À-} +i[ÐA^{À-À-},A^{--}] \cr
& -i{1\over (d^{+À+})^3} (Ðd^{À+})^4 [ÐA^{À-À-},A^{+À-}]
	+{1\over (d^{+À+})^4}d^+{}_a (Ðd^{À+})^4 [ÐA^{À-a},A^{+À-}] \cr} $$
Using $d^+{}_a d^{--} ÐA^{À-À-} = d^-{}_a ÐA^{À-À-}$,
$$ \li{ d^-{}_a ÐA^{À-À-} = ¼& -i{1\over d^{+À+}}d_{ab}ÐA^{À+b} +i[ÐA^{À-À-},A^-{}_a] \cr
	& +{1\over (d^{+À+})^2} \leftÓ -(Ðd^{À+})^3{}_a[ÐA^{À-À-},A^{+À-}] 
	+ ü·_{abcd}Ðd^{À+b} Ðd^{À+c} [ÐA^{À-d},A^{+À-}] \rightÕ \cr} $$

There are now several ways to solve for the field equation $d^{-À-}A^{--}=iHA^{--}$, which is the lightcone analog of the nonrelativistic Schr¬odinger equation in terms of the ``Hamiltonian" $H$:  One way, as outlined for the bosonic case, is to first solve for $A^{-À-}$, from
$$ [ Ñá^{À-À-},á^{-À+} ] = - á^{-À-}¼¼Ü¼¼
	A^{-À-} = d^{-À+}ÐA^{À-À-} + d^{+À-}A^{--} - d^{+À+}Ðd^{À-À-}A^{--} +i[A^{-À+},ÐA^{À-À-}] $$
(another case of $d^{i-}A^{j-}$ from the bosonic section), use the above for $Ðd^{À-À-}A^{--}$ to give
$$ \li{ A^{-À-} = ¼& ( d^{-À+}ÐA^{À-À-} + d^{+À-}A^{--}
	-\f14 ·^{abcd}d_{ab}A_{cd}{}^{-À-} ) +i[A^{-À+},ÐA^{À-À-}] \cr
	& +i{1\over (d^{+À+})^2} (Ðd^{À+})^4 [ÐA^{À-À-},A^{+À-}]
	- {1\over (d^{+À+})^3} d^+{}_a (Ðd^{À+})^4 [ÐA^{À-a},A^{+À-}] \cr} $$
and then perform similar manipulations for $d^{--}$ on $A^{-À-}$ in
$$ [ á^{--},á^{-À-}] = 0âÜâd^{-À-}A^{--} = d^{--}A^{-À-} + i[A^{--},A^{-À-}] $$
An equivalent way is to expand $[d^{--},Ðd^{À-À-}]A^{--}=0$.  

Another way is to evaluate $Ód^-{}_a,Ðd^{À-b}Õ=-i¶_a^b d^{-À-}$ on $A^{--}$.
We'll do this in a simpler way:  The $d^-$ field equation follows directly from varying the action, so we'll evaluate the action directly as an anticommutator, as described below.

ÜSymmetries

An alternative and commonly used way to derive the analog of the above expressions for symmetry generators is to use canonical transformations, i.e., second-quantized commutators (see [30] for supersymmetry).  We'll review some of the features of this approach here, and show how covariant derivatives can be used to shortcut some of the initial steps.  After solving the above equations, symmetry generators, but not all covariant derivatives, can be expressed as such operators, since the prepotential is then defined only on the reduced (lightcone) chiral superspace.  Since we'll concentrate on deriving the action,  we'll reduce vector derivatives to just partial derivatives.

There is a term $àüÄ »^+ »^- Ä$ in any lightcone action.  (The sign depends on the integration measure:  It's $+$ for a bosonic scalar, but $-$ for our chiral superfield.  This is clear from a component expansion in evaluating $Çd^4 ÏÊA^{--}õA^{--}$, using $Çd^4 Ï=(d^+)^4$.)  Thus $¦»^+ Ä$ is the ``momentum" for $Ä$, so a general (super)field will satisfy the canonical equal-time commutation relations
$$ [Ä(1),Ä(2)] = ài{1\over »^+}¶(1-2) $$
where the $¶$-function is in all lightcone chiral superspace coordinates except the lightcone time $x^+$.  (We're really working in the Schr¬odinger picture, so there is no time.)  Symmetry generators (including the Hamiltonian) are then functionals of the field $\G[Ä]$ (quadratic in the free case), acting as
$$ ¶Ä = [\G,Ä] = ¦i{1\over »^+}{¶\G\over ¶Ä} $$

Conversely, $\G$ can be found from $¶Ä$ by integrating $¶\G/¶Ä=ài»^+ ¶Ä$.  
For example, in the free case the second-quantized linear symmetry operators $\G$ follow from first-quantized linear ones $G$ as
$$ \G = àiÇüÄ »^+ GÄâÜâ[\G,Ä] = GÄ $$
$$ [G_1,G_2Õ = G_3âÜâ[\G_1,\G_2Õ = -\G_3 $$
where ``$Ç$" integrates over all the coset coordinates except the lightcone time $x^+$.

At the classical level, we then have the Poisson bracket (with quantum mechanical normalization)
$$ [\G_1,\G_2Õ = àiÇ{¶\G_1\over ¶Ä}{1\over »^+}{¶\G_2\over ¶Ä} = 
	àiÇ(¶_1 Ä)»^+(¶_2 Ä) $$
(again integrated over everything except $x^+$).  Bose symmetrization, together with the antisymmetry of the $1/»^+$, then gives the (graded) commutator of $¶_1$ and $¶_2$.
In particular, for supersymmetry we have $Ó©^- q,©^- qÕ=-i©^- p^-$, so the Hamiltonian $-ip^-=ip_+$ can be derived as quadratic in the supersymmetry variations.  

We modify this approach here by first using the solution to the equations for the covariant spinor derivatives to find the supersymmetry generators.  This step is trivial, since the interaction terms in the covariant spinor derivatives $d$ and supersymmetry transformations are identical:
Comparing the expressions for the covariant derivatives $d$ and symmetry generators, we have for those with corrections from interactions (for ``$¶$" in $¶A^{--}$):
$$ \li{ J^{i-} & = (x^i d^- - x^- »^i) - Ï ©^i ©^- (i ü ©^i »^i Ï + d) + d^{i-} \cr
	©^- q & = -i ©^i »^i (©^- Ï) + ©^- d \cr
	p^- & = d^- \cr} $$
where we've used the fact that $q$ and the spinor derivative $d$ have the same $»/»Ï$ term and opposite $Ïp$ term, and applied the gauge condition $©^+ Ï=0$. (There are generally also unitary transformations to simplify the lightcone representation, involving $Ï$ and $»_x$ but not $»_Ï$, which therefore leave the above relation between $q$ and $d$ unchanged.)  The interactions appear only upon solving the equations of motion for the partial derivatives with respect to the non-lightcone coordinates $y^{i+}$, $©^+ Ï$, and $x^+$.  (We solve for the derivatives, then set the corresponding coordinates to vanish.)  In the way we have expressed the generators above, all these partial derivatives are contained in covariant derivatives.  Thus all interaction terms in the generators can be read directly from the covariant derivatives $d^{i-}$, $©^- d$, and $d^-$.  

At least for the case of supersymmetry, it'll be useful to go one step further and make a similar replacement for $©^- Ï$ by comparing $©^+q$ and $©^+d$:  we then find
$$ ©^- Ï = i{1\over »^+}©^-(©^+ q -©^+d) $$
$$ Üâ©^- q = {1\over »^+}(©^i »^i)©^-(©^+ q -©^+ d) +©^- d $$
The explicit $©^+ q$ term is the usual free part of $©^- q$ as found from solving $Öpq=0$; similarly, the $d$ terms are the interacting $©^- d$ minus the free part found from $Öpd=0$.  In particular, in euphoric notation we have
$$ \li{ q^-{}_a & = 
	{1\over »^{+À+}}[(q^+{}_a -d^+{}_a)»^{-À+} +i(Ðq^{À+b} -Ðd^{À+b})»_{ba}] +d^-{}_a, \cr
	Ðq^{À-a} & = {1\over »^{+À+}}[(Ðq^{À+a} -Ðd^{À+a})»^{+À-} 
		+i(q^+{}_b -d^+{}_b)ü·^{bacd}»_{cd}] +Ðd^{À-a} \cr} $$

In our case the role of the fundamental field $Ä$ is played by $A^{--}$.  Thus in particular we can use the above results for $d^-{}_a A^{--}$ and $Ðd^{À-a} A^{--}$ to find the corresponding transformations $©^-q$, and express the generator for $p^-$ by multiplying these supersymmetry variations.  (The Lorentz generators are also straightforward.  An interesting case is $J^{--}$:  The $d^{--}$ terms serve only to cancel terms generated from the rest by Bose symmetrization.)
Substituting our solution for $©^-d$ into $©^-q$, we have
$$ \li{ q^-{}_a A^{--} & = {1\over »^{+À+}}Ó(»^{-À+}q^+{}_a -i»_{ab}Ðq^{À+b}) A^{--}
	-i[q^+{}_a A^{--},»^{+À+}A^{--}]Õ \cr
Ðq^{À-a}A^{--} & = {1\over »^{+À+}}\leftÓ (»^{+À-}Ðq^{À+a} -iü·^{abcd}»_{bc}q^+{}_d) A^{--}
	-i{(Ðd^{À+})^4\over (»^{+À+})^2} [Ðq^{À+a}ÐA^{À-À-},»^{+À+}ÐA^{À-À-}] \rightÕ \cr} $$
The $Ï$ terms in the commutators drop out automatically, allowing us to also replace those $d$'s with $q$'s.  (We can also use the complex conjugate of $d^-{}_a ÐA^{À-À-}$ to write $Ðq^{À-a}A^{--}$ directly in terms of just $A^{--}$ and not $ÐA^{À-À-}$, but with more terms.)  We've written the transformations in this way to make explicit the fact that they preserve the (anti)chirality of $A^{--}$ ($ÐA^{À-À-}$).  (We could use the chiral representation, where $q^+=»^+$, or the antichiral one, where $Ðq^{À+}=л^{À+}$.)

Showing the product $Ç(q^-{}_a A^{--})»^{+À+}(Ðq^{À-b}A^{--})$ is proportional to $¶_a^b$ is still complicated, but the trace is easy to evaluate.  The calculation then proceeds as follows:  
In the quadratic (free) terms, only those with an even number of derivatives survive Bose symmetrization, forcing the $q$'s to contribute only as their anticommutators ($qÐq£üÓq,ÐqÕ$), giving the usual result.  In the cubic terms there is a similar Bose symmetrization identity
$$ trÇ(»_A Ä)[»_B Ä,»_C ÄÕ = 0 $$
since the antisymmetry of the structure constants (commutator) forces (graded) antisymmetry in the indices of the derivatives, but integration by parts gives two derivatives on a single field, which must be symmetric.  This can be used in the case with both $q$ and $Ðq$ since, e.g., $ÐqA=(Ðq-Ðd)A=iÏ»^{+À+}A$, again forcing an anticommutator.
In the interaction terms involving $(Ðd^{À+})^4$, we can use it to convert the $Çd^4 Ï$ into $Çd^8 Ï$.  In the corresponding cubic terms, we can then do the $Çd^4 Ï$ on the $A^{--}$ to leave an $Çd^4 ÐÏ$ on three $ÐA^{À-À-}$'s.

The result is then, abbreviating $Ä=A^{--}$, $ÐÄ=ÐA^{À-À-}$, $»^+=»^{+À+}$, $»_a=»^+{}_a$, and $л^a=л^{À+a}$ (and again we can replace $q_a$ with $»_a$ or $d_a$ inside the commutators):
$$ S = {1\over g^2}ÊtrÇd^4 x\left[Çd^4 Ï \left( -üÄ»^+ »^- Ä 
	+\H_2 +\H_3 \right)  +Çd^4ÐÏÊÑ{\H}_3 +Çd^4ÏÊd^4ÐÏÊ\H_4\right] $$
$$ \li{ \H_2 & = üÄ(»^{+À-}»^{-À+} -\f18·^{abcd}»_{ab}»_{cd})Ä \cr
	\H_3 & = -iüÄ[»^+Ä,»^{+À-}Ä] 
		-i\f18·^{abcd}(»_{ab}»_c Ä){1\over »^+}[»^+Ä,»_d Ä]  \cr
	Ñ{\H}_3 & = -iüÐÄ[»^+ÐÄ,»^{-À+}ÐÄ] 
		-i\f14(»_{ab}л^a ÐÄ){1\over »^+}[»^+ÐÄ,л^b ÐÄ] \cr
	\H_4 & =- i\f14[»^+Ä,»_a Ä]{1\over (»^+)^3}[»^+ÐÄ,л^a ÐÄ] \cr} $$

Û0 Conclusions

We have introduced new coordinates for supersymmetric theories by a combination of covariant coset and dimensional reduction, which are formally ``left and right coset" but are distinguished physically by their relation to symmetry vs.¼covariant derivatives (as determined by unitarity) and free vs.¼coupled.  The extension is suggested by regarding equally the spinor and antisymmetric tensor field strengths of super Yang-Mills (or their ``squares" in supergravity), as gauge fields before dimensional reduction.  This implies similar treatment for supersymmetry coordinates (which introduce spin) compared to Lorentz coordinates (which introduce superspin), which are dual to these field strengths.  Related remarks apply to superstrings, where the spinor field strength was already known to appear so.

The results here suggest several avenues of extension:  (1) We have shown how the covariant (in all ways) formulation is always the best starting point, even for lightcone applications.  The field theory equations are equivalent to the first-class constraints of first-quantization.  We applied the analysis only to the super Yang-Mills case in generality; (non-selfdual) supergravity would also be interesting:  The selfdual case already confirms that the chiral prepotential $h^{----}$ corresponds to the square of the super Yang-Mills $A^{--}$.  

(2) Lorentz coordinates and the related R-symmetry coordinates have proven advantageous previously in projective and harmonic superspace approaches.  Perhaps similar methods will solve the problem of extending covariant formulations of maximally supersymmetric theories off shell.  The prepotentials we have found are of the appropriate (engineering) dimension for such theories.  In particular, the new dual Lorentz coordinates are necessary for this result in the supergravity case.

(3) The new affine Lie algebra for superstrings implies modifications of first-class constraints for superstrings, including generalizations of previous covariant approaches such as pure spinors.  Such considerations are particularly natural for anti de Sitter strings, where the Lorentz algebra is already required for closing the algebra of ``translations".

ÜAcknowledgment

This work is supported in part by National Science Foundation Grant No.¼PHY-0969739.

\refs

£1  
  P.A.M. Dirac,
  ÓAnnals Math.Õ É37 (1936) 429;\\
  H.A. Kastrup,
  ÓPhys. Rev.Õ É150 (1966) 1183;\\
  G. Mack and A. Salam,
  ÓAnnals Phys.Õ  É53 (1969) 174;\\
  S.L. Adler,
  \PRD 6 (1972) 3445;\\
  R. Marnelius and B.E.W. Nilsson,
  \PRD 22 (1980) 830.

£2  
  A. Karlhede, U. Lindstr¬om, and M. Ro×cek,
  \PL 147B (1984) 297;\\
  U. Lindstr¬om and M. Ro×cek,
  ÓCommun. Math. Phys.Õ  É115 (1988)  21,
É128 (1990) 191.

£3  
  A. Galperin, E. Ivanov, S. Kalitsyn, V. Ogievetsky and E. Sokatchev,
  ÓClass. Quant. Grav.Õ  É1 (1984) 469.

£4 
  A. Galperin, E. Ivanov, V. Ogievetsky and E. Sokatchev,
  ÓJETP Lett.Õ  É40 (1984) 912;\\
  B.M. Zupnik,
  ÓTheor. Math. Phys.Õ É69 (1986) 1101.

£5 
  E. Sokatchev,
  \PL 169B (1986) 209.

£6  
  M. Hatsuda, Y.-t. Huang, and W. Siegel,
  ÓJHEPÕ É0904 (2009) 058
  \xxxlink{0812.4569} [hep-th].

£7 
  P.S. Howe,
  \PL 258B (1991) 141,
  É259B (1991) 511,
  É273B (1991) 90;\\
  N. Berkovits,
  ÓJHEPÕ É0004 (2000) 018
  \xxxlink{hep-th/0001035}.

£8 W. Siegel, Covariant approach to superstrings, ã Symposium on anomalies, geometry and topology, eds. W.A. Bardeen and A.R. White (World Scientific, Singapore, 1985) p. 348.

£9 
W. Siegel,
  \NP 263 (1986) 93.

£10 
  K. Lee and W. Siegel,
  ÓJHEPÕ É0606 (2006) 046
  \xxxlink{hep-th/0603218}.

£11 
  M.B. Green,
  \PL 223B (1989) 157.

£12  
  W. Siegel,
  \PRD 50 (1994) 2799
  \xxxlink{hep-th/9403144}.

£13 
  M. Sakaguchi,
  \PRD 59 (1999) 046007
  \xxxlink{hep-th/9809113}.

£14 
  W. Siegel,
  \PRD 48 (1993) 2826
  \xxxlink{hep-th/9305073};
  Manifest duality in low-energy superstrings,
  ã Strings '93, eds. M.B. Halpern, G. Rivlis, and A. Sevrin (World Scientific, Singapore, 1993) p. 353
  \xxxlink{hep-th/9308133}.

£15 
  K. Lee and W. Siegel,
  ÓJHEPÕ É0508 (2005) 102
  \xxxlink{hep-th/0506198}.

£16 
  S.W. MacDowell and F. Mansouri,
  \PR 38 (1977) 739.

£17  
  W. Siegel,
Free field equations for everything, 
ã Superstrings, Cosmology, Composite Structures, 
eds. S.J. Gates, Jr. and R.N. Mohapatra (World Scientific, Singapore, 1987) p. 13;
  \PL 203B (1988)  79,
  ÓMod. Phys. Lett.Õ  ÉA5 (1990)  2767.

£18  
  N. Berkovits,
  ÓJHEPÕ É0409 (2004) 047
  \xxxlink{hep-th/0406055};\\
  I. Oda and M. Tonin,
  \NP 727 (2005) 176
  \xxxlink{hep-th/0505277}.

£19  
  A.N. Leznov,
  ÓTheor. Math. Phys.Õ É73 (1988) 1233;\\
  A.N. Leznov and M.A. Mukhtarov,
  ÓJ. Math. Phys.Õ É28 (1987) 2574;\\
  A. Parkes,
  \PL 286B (1992) 265
  \xxxlink{hep-th/9203074}.

£20  
  W. Siegel,
  \PRD 47 (1993) 2504
  \xxxlink{hep-th/9207043}.

£21 
  C. Devchand and V. Ogievetsky,
  \NP 414 (1994) 763
  \xxxlink{hep-th/9306163};\\
  E. Sokatchev,
  \PRD 53 (1996) 2062
  \xxxlink{hep-th/9509099};\\
  S. Karnas and S.V. Ketov,
  \NP 526 (1998) 597
  \xxxlink{hep-th/9712151}.

£22  
  G. Chalmers and W. Siegel,
  \PRD 54 (1996) 7628
  \xxxlink{hep-th/9606061}.

£23 
  J.F. Pleba«nski,
  ÓJ. Math. Phys.Õ É16 (1975) 2395.

£24 
  W. Siegel,
  \PRD 53 (1996)  3324
  \xxxlink{hep-th/9510150}.

£25 
  A.S. Galperin, N.A. Ky, and E. Sokatchev,
  ÓClass. Quant. Grav.Õ É4 (1987) 1235;\\
  A.S. Galperin, E.A. Ivanov, V.I. Ogievetsky, and E. Sokatchev,
  ÓClass. Quant. Grav.Õ É4 (1987) 1255.

£26 
  M.T. Grisaru and W. Siegel,
  \NP 201 (1982) 292.

£27  
  S. Weinberg,
  ÓPhys. Rev.Õ  É150 (1966) 1313;\\
  J.B. Kogut and D.E. Soper,
  \PRD 1 (1970) 2901.

£28  
  J. Scherk and J.H. Schwarz,
  ÓGen. Rel. Grav.Õ É6 (1975)  537;\\
M. Kaku, \NP 91 (1975) 99;\\
  M. Goroff and J.H. Schwarz,
  \PL 127B (1983)  61.

£29  
  S. Mandelstam,
  \NP 213 (1983) 149;\\
  L. Brink, O. Lindgren, and B.E.W. Nilsson,
  \NP 212 (1983)  401.

£30  
  A.K.H. Bengtsson, I. Bengtsson, and L. Brink,
  \NP 227 (1983) 41;\\
  S. Ananth, L. Brink and P. Ramond,
  ÓJHEPÕ É0407 (2004) 082
  \xxxlink{hep-th/0405150};\\
  S. Ananth, L. Brink, S.S. Kim, and P. Ramond,
  \NP 722 (2005) 166\\
  \xxxlink{hep-th/0505234}.

£31  
  R.R. Metsaev,
  \PRD 71 (2005) 085017
  \xxxlink{hep-th/0410239};\\
  S. Ananth, L. Brink, and P. Ramond,
  ÓJHEPÕ É0505 (2005) 003
  \xxxlink{hep-th/0501079};\\
  L. Brink, S.S. Kim, and P. Ramond,
  ÓJHEPÕ É0806 (2008) 034
  [ÓAIP Conf. Proc.Õ É1078 (2009) 447]
  \xxxlink{0801.2993} [hep-th].

£32  
  M.F. Sohnius,
  \NP 136 (1978) 461;\\
  E. Witten,
  \PL 77B (1978) 394;\\
  L. Brink and P.S. Howe,
  \PL 88B (1979)  268;\\
  E. Cremmer and S. Ferrara,
  \PL 91B (1980) 61;\\
  L. Brink and P.S. Howe,
  \PL 91B (1980) 384;\\
  W. Siegel,
  \NP 177 (1981) 325.

£33 
  W. Siegel,
  \PL 80B (1979)  220.

£34 
  E. Witten,
  \NP 266 (1986)  245.

\bye